\documentclass[sigconf]{acmart}

\usepackage{hyperref}
\usepackage{latexsym}
\usepackage{mathrsfs}
\usepackage{multirow}
\usepackage{amsmath}
\usepackage{amsfonts}
\usepackage{subfigure}
\usepackage{graphicx}
\usepackage{tabularx}
\usepackage[boxed,ruled,lined]{algorithm2e}
\usepackage{algorithmic}
\usepackage{comment}
\usepackage{balance}
\newtheorem{theorem}{Theorem}

\AtBeginDocument{%
  \providecommand\BibTeX{{%
    \normalfont B\kern-0.5em{\scshape i\kern-0.25em b}\kern-0.8em\TeX}}}

\copyrightyear{2022} 
\acmYear{2022} 
\setcopyright{acmcopyright}\acmConference[CIKM '22]{Proceedings of the 31st ACM International Conference on Information and Knowledge Management}{October 17--21, 2022}{Atlanta, GA, USA}
\acmBooktitle{Proceedings of the 31st ACM International Conference on Information and Knowledge Management (CIKM '22), October 17--21, 2022, Atlanta, GA, USA}
\acmPrice{15.00}
\acmDOI{10.1145/3511808.3557299}
\acmISBN{978-1-4503-9236-5/22/10}

\begin{document}
\begin{sloppy}
%\orphanpenatly = 10 
%\widowpenaty = 1
%%
%% The "title" command has an optional parameter,
%% allowing the author to define a "short title" to be used in page headers.
%\title{User Preference Prediction in Multi-round Conversational Recommendation with Feedback Revision}
\title{Dually Enhanced Propensity Score Estimation \mbox{in Sequential Recommendation}}

\author{Chen Xu}

%\authornotemark[1]
\affiliation{%
  \institution{Gaoling School of Artificial Intelligence}
  \country{Renmin University of China}
  \\xc\_chen@ruc.edu.cn
}

% \footnote{* corresponding author.}
\author{Jun Xu}
\authornote{Jun Xu is the corresponding author. Work partially done at Beijing Key Laboratory of Big Data Management and Analysis Methods.}
\affiliation{%
  \institution{Gaoling School of Artificial Intelligence}
  \country{Renmin University of China}
  \\junxu@ruc.edu.cn
}

\author{Xu Chen}
\affiliation{%
  \institution{Gaoling School of Artificial Intelligence}
  \country{Renmin University of China}
  \\xu.chen@ruc.edu.cn
}

\author{Zhenghua Dong}
%\authornotemark[1]
\affiliation{%
  %\institution{Huawei Noah's Ark Lab}
  \country{Huawei Noah's Ark Lab}
  \\dongzhenhua@huawei.com
}

\author{Ji-Rong Wen}
\affiliation{%
  \institution{\mbox{Gaoling School of Artificial Intelligence}}
  \country{Renmin University of China}
  \\jrwen@ruc.edu.cn
}
\renewcommand{\shortauthors}{Chen Xu et al.} 
%% No italics and no comma 
%% If needed use a foot or author note to identify equal contribution
\begin{abstract}
Sequential recommender systems train their models based on a large amount of implicit user feedback data and may be subject to biases when users are systematically under/over-exposed to certain items. Unbiased learning based on inverse propensity scores (IPS), which estimate the probability of observing a user-item pair given the historical information, has been proposed to address the issue. In these methods, propensity score estimation is usually limited to the view of item, that is, treating the feedback data as sequences of items that interacted with the users. However, the feedback data can also be treated from the view of user, as the sequences of users that interact with the items. Moreover, the two views can jointly enhance the propensity score estimation. Inspired by the observation, we propose to estimate the propensity scores from the views of user and item, called Dually Enhanced Propensity Score Estimation (DEPS). Specifically, given a target user-item pair and the corresponding item and user interaction sequences, DEPS firstly constructs a time-aware causal graph to represent the user-item observational probability. According to the graph, two complementary propensity scores are estimated from the views of item and user, respectively, based on the same set of user feedback data. Finally, two transformers are designed to make the final preference prediction. Theoretical analysis showed the unbiasedness and variance of DEPS. Experimental results on three publicly available and an industrial datasets demonstrated that DEPS can significantly outperform the state-of-the-art baselines. %The empirical analysis also showed that DEPS can be used as a model-agnostic de-biasing framework under which other sequential recommendation models can also be improved. make use of the two propensity scores and
\end{abstract}



\maketitle
 
\section{Introduction}\label{sec:intro}
Sequential recommendation~\cite{bert4rec,imporvedRec4gru,stamp} has attracted increasing attention from both industry and academic communities. 
Basically, the key advantage of sequential recommender models lies in the explicit modeling of item chronological correlations.
To capture such information accurately, recent years have witnessed lots of efforts based on either Markov chains or recurrent neural networks. While these models have achieved remarkable successes, the observed item correlations can be skewed due to the exposure or selection bias~\cite{Survy:Unbias-Rec}.
As exampled in Figure~\ref{intro}(a), given a user behavior sequence, the observed next item is a coffeepot cleaner. 
By building models based on the observational data, one can learn the correlations between cleaner and coffeepot.
However, from the user preference perspective, the next item can also be the ink-boxes. But the model has no opportunities to capture the correlations between printer and ink-box because they are not recommended and observed in the data. The bias makes the recommendation less effective, especially when testing environment is more related with the office products.

\begin{figure}
    \centering
    \includegraphics[width = 0.4\textwidth]{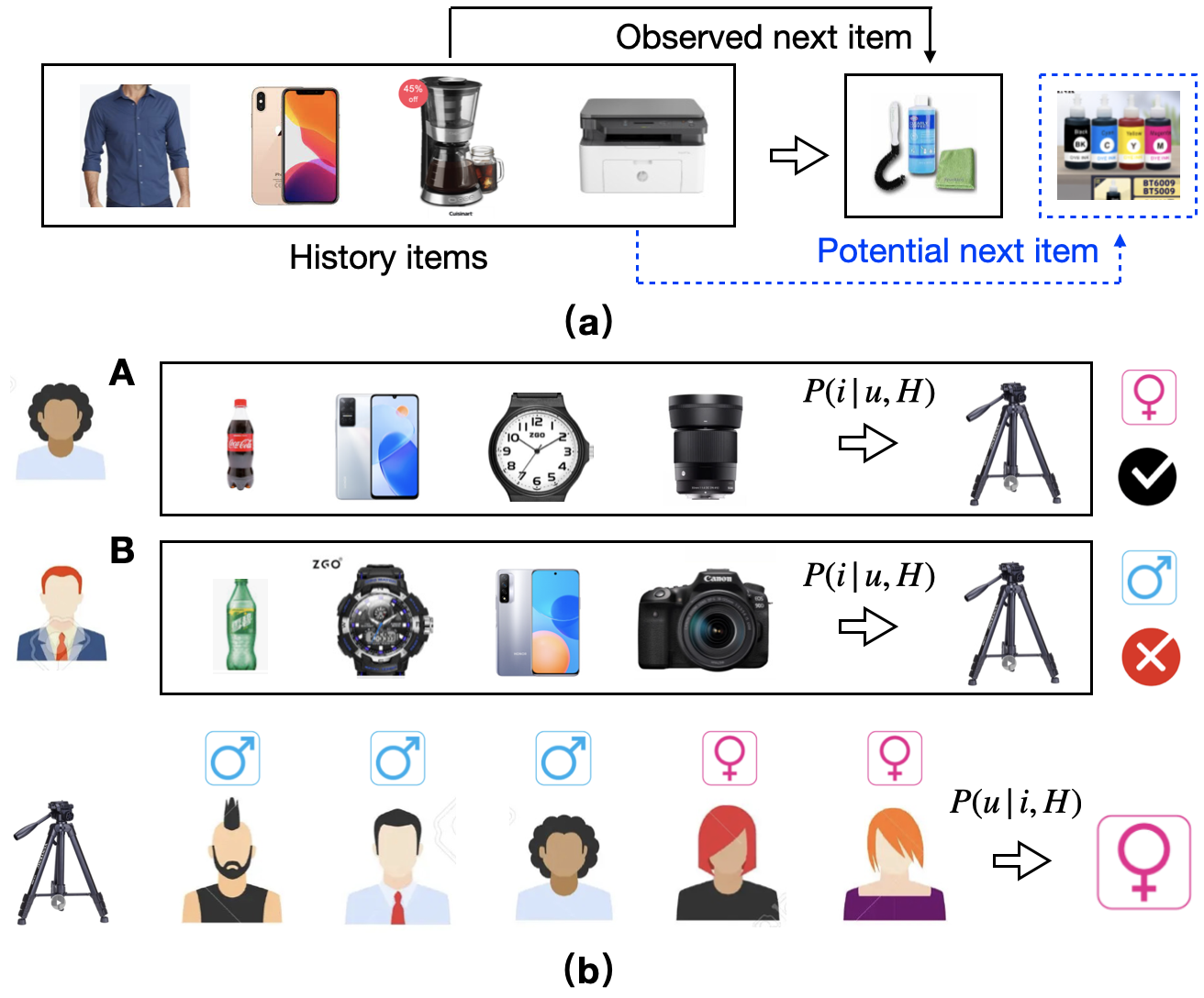}
    \caption{
    (a) Motivating example of unbiased sequential recommendation.
    (b) A toy example on the complementary roles of the user prediction problem for existing IPS methods.} 
    \label{intro}
\end{figure}

In order to alleviate the above problem, previous models are mostly based on the technique of inverse propensity score (IPS)~\cite{Schnabel:RecAsTreat}, where if a training sample is more likely to appear in the dataset, then it should have lower weight in the optimization process.
In this research line, the key is to accurately approximate the probability of observing a user-item pair $(u, i)$ given the historical information $H$, i.e., $P(u,i|H)$.
To this end, previous methods usually decompose $P(u,i|H)$ as $P(i|u,H)P(u|H)$, and focus on parameterizing $P(i|u,H)$ (i.e., estimating $P(u,i|H)$ from view of item~\cite{wang2022unbiased}) to predict which item the user will interact in the next given the previous items. One reason is that estimating propensity scores from the view of item matches well with the online process of sequential recommendation: the users come to the system randomly and the system aims to provide the recommended items immediately.  %and predicting item scores given the user. 

%%\footnote{Note that, in online process, the sequential recommender system focused on predicting the scores of item are recommended given the user and the historical information, i.e. $r(i|u,H)$. However, the key for estimating propensity score is to accurately approximate the probability of observing a user-item pair $(u, i)$ given the historical information $H$, i.e., $P(u,i|H)$. }

While these methods are effective, we argue that the probability of observing a user-item pair can also be considered from a dual perspective, that is, for an item, predicting the next interaction user given the ones who have previously interacted with it.
In principle, this is equal to decomposing $P(u,i|H)$ in another manner by $P(u|i,H)P(i|H)$, where $P(u|i,H)$ exactly aims to predict the user given an item and the history users (i.e. estimating $P(u,i|H)$ from view of user).
Intuitively, for the same item, if two users interact with it for a short time, they should share some similarities at that time.
As a result, the previous users may provide useful signals~\cite{fan2021continuous} for predicting the next user and the observation of the user-item pair.
We believe such a user-oriented method can provide complementary information to the previous item-oriented models.

For example, in Figure~\ref{intro}(b),
from the item prediction perspective, the tripod can be observed as the next item for both sequences A and B, since the historical information is similar.
However, from the user prediction perspective, we may infer that sequence A should be more likely to be observed, since recently, the tripod is more frequently interacted with by female users, for example, due to the reasons like the promotion sales for the Women's Day.
This example suggests that the temporal user correlation signals may well compensate for traditional item-oriented IPS methods, which should be taken seriously for debiasing sequential recommender.

%Motivated by the above analysis, 
This paper proposes to build an unbiased sequential recommender model with dually enhanced IPS estimation (called DEPS).
The major challenges lie in three aspects:
% to begin with, the user preferences in real-world scenarios can be quite complex, and how to capture their correlations can be not easy, especially, since we have to consider the chronological information.
% Then, both the item- and user-oriented IPS methods are useful. How to combine them is still not clear.
To begin with, the item- and user-oriented IPS are useful, how do we estimate them from the same set of the user feedback data? Secondly, how to combine them is still not clear, especially, since we have to consider the chronological information.
At last, how to theoretically ensure that the proposed objective is still unbiased also needs our careful design.

To solve these challenges, we use two GRUs to estimate the propensity scores, one from the view of item and another from the view of user. Also, to make our model DEPS practical, two transformers are used to make the final recommendation, one encodes the historical interacted item sequence of the target user, and the other encodes the user sequence that interacted with the target item. The encoded sequences' embeddings, as well as the target item and user embeddings, are jointly used to predict the final recommendation score. Moreover, a two-stage learning procedure is designed to estimate the parameters from the modules of propensity score estimation and the item recommendation.

Major contributions of this paper can be concluded as follows:

(1) We highlighted the importance of propensity score estimation from two views and proposed a dually enhanced IPS method for debiasing sequential recommender models.

(2) To achieve the above idea, we implement a double GRU architecture to consider both user- and item-oriented IPS estimation, and theoretically proof the unbiasedness of our objective.

(3) We conduct extensive experiments to demonstrate the effectiveness of our model by comparing it with the state-of-the-art methods based on three publicly available benchmarks and an industrial-scale commercial dataset.

\section{Related Work}
A lot of research efforts have been made to develop models for sequential recommendation~\cite{bert4rec,DIN,BST,dien,imporvedRec4gru,stamp,lightsans}. Compared to traditional recommendation~\cite{DMF,BPR}, sequential recommendation tries to capture the item chronological correlations. Models based on either Markov chains or recurrent neural networks have been proposed. For example, GRU4Rec+~\cite{imporvedRec4gru} introduces an RNN to encode the historical item sequences as the user preference. BERT4Rec~\cite{bert4rec} proposes an attention-based way~\cite{attention} to model user behavior sequences practically. BST~\cite{BST} utilizes the transformer~\cite{attention} to capture the user preference from the interaction sequences. LightSANs~\cite{lightsans} introduce a low-rank decomposed self-attention to the model context of the item. As for model training, S3-Rec~\cite{S3-rec} incorporates self-supervised and adapts the Pre-train/fine-tune paradigm.

Modern recommender systems have to face variant biases, including selection bias~\cite{marlin12:SeB4CF}, position bias~\cite{zheng20:disentangling,collins18:PB4DigitRe}, popularity bias~\cite{zhang:PopBias4RS}, and exposure bias~\cite{liu20:EB4KDF,saito20:UIR,chen19:samwalker, chen20:FAWMF,Abdollahpouri:Multi-EB}.
Biases usually happen on multi sides~\cite{Abdollahpouri:Multi-EB, Survy:Unbias-Rec}. For example, item exposure is affected by both the user's previous behaviors~\cite{liu20:EB4KDF,saito20:UIR} and the user's background~\cite{chen:18MUE4SN,chen19:samwalker, chen20:FAWMF}.~\citet{wang:CI4SRS,zhang:Causerec,wang2022unbiased} pointed out that sequential scenarios are different and more studies are needed.

One common way to remedy the bias is through inverse propensity score (IPS)~\cite{Schnabel:RecAsTreat}.~\citet{hu:08CF4Implicit,devooght:15dynamicMF} used the prior experience as propensity score to uniformly re-weight the samples. UIR~\cite{saito20:UIR} and UBPR~\cite{ubpr} propose to utilize the latent probabilistic model to estimate propensity score.~\citet{agarwal2019estimating, fang2019intervention} utilized the intervention harvesting to learn the propensity.~\citet{qin2020attribute, joachims2017unbiased} learns propensity model with EM algorithms.USR~\cite{wang2022unbiased} proposed a network to estimate propensity scores from the view of item in the sequential recommendation.~\cite{chen:18MUE4SN,chen19:samwalker} pointed out that it is useful to carefully consider the user's perspective when estimating propensity.

%Many researches estimated propensity score from view of item (i.e. assumed that propensity score is user independent~\cite{Yang:18OfflineEB}). 
%For example,~\citet{hu:08CF4Implicit,devooght:15dynamicMF} used the prior experience to uniformly re-weight the samples,  ~\citet{wang2022unbiased} proposed to estimate the propensity score from view of item in the sequential recommendation. 

%To help the users to remedy the exposure or self-selection bias, this paper studies the problem of propensity estimation in the sequential recommendation. 

\section{Problem Formulation}
%This section formulates the problem of sequential recommendation and analyzes the bias in the sequential recommendation.% and unbiased learning for the recommendation. % the problem with the help from time-based causal graph. 

\begin{figure*}[h]
    \centering
    \includegraphics[width=0.88\textwidth]{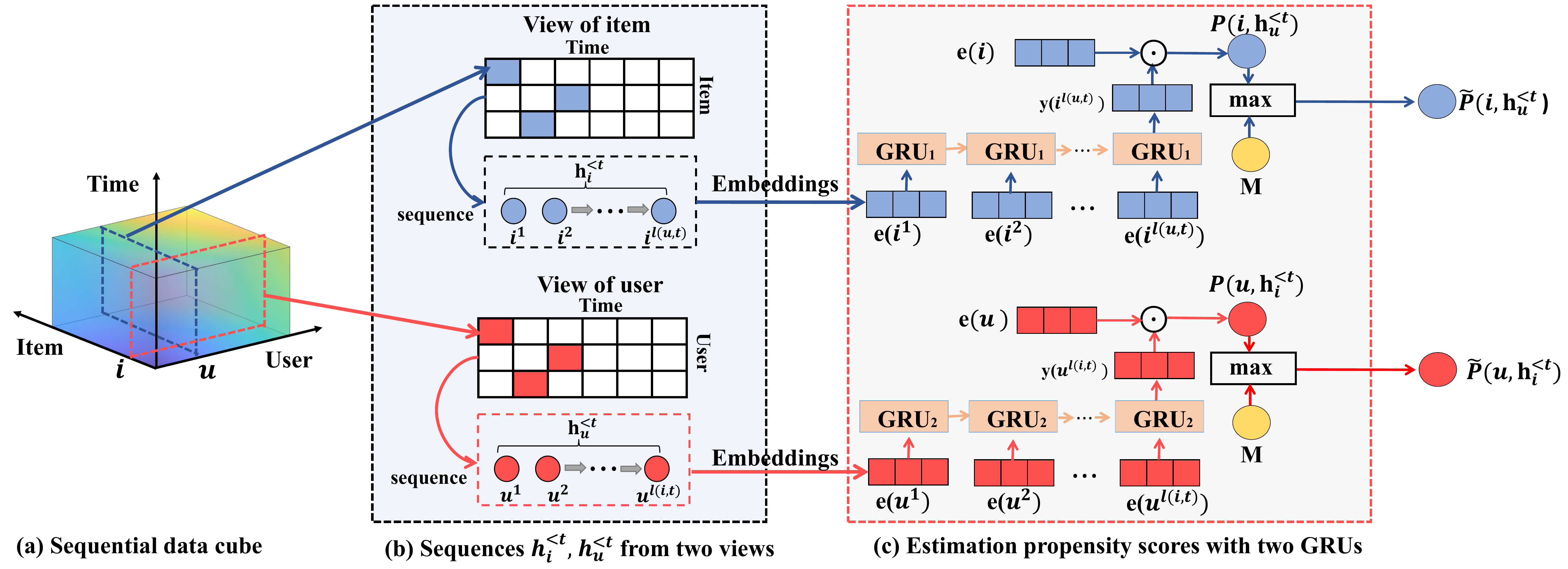}
    \caption{Propensity score estimation in sequential recommendation. (a) Representing historical user-item interactions as a data cube, where item and user can be placed on the axis in any order; (b) Construction of two interaction sequences correspond to a target tuple $(u,i,t)$, from the views of item and user, respectively; (c) Using two GRUs to estimate the propensity scores. }
    \label{fig:IPS_estimation}
    %\vspace{-0.4cm}
\end{figure*}
\subsection{Sequential Recommendation}
Suppose that a sequential recommender system manages a set of user-item historical interactions $\mathcal{D} = \{(u,i,c_t)\}$
were each tuple $(u, i, c_t)$ records that at time stamp $t$, a user $u\in\mathcal{U}$ accessed the system and interacted with an item $i\in\mathcal{I}$, and the user's feedback is $c_t\in\{0, 1\}$, where $\mathcal{U}$ and $\mathcal{I}$ respectively denote the set of users and items in the system, and $c_t=1$ means that the user $u$ clicked the item $i$ and 0 otherwise at time $t$. Moreover, the context information of $(u,i)$ (e.g. user profile and item attribute) collected from the system is often represented as real-valued vectors (embeddings) $\mathbf{e}(u),\mathbf{e}(i) \in \mathbb{R}^d$, where $d$ denotes the dimensions of the embeddings.

At a specific time $t$ and given a target user-item pair $(u, i)$, two types interaction sequences can be derived from $\mathcal{D}$: (1) the sequence of items that the user $u$ previously interacted before time $t$: $\mathbf{h}^{<t}_u = [i^1,i^2,\cdots,i^{l(u,t)}]$, where $l(u,t)$ denotes the number of items the user $u$ interacted before time $t$; (2) the sequence of the users that item $i$ was previously interacted with before time $t$: $\mathbf{h}^{<t}_i = [u^1,u^2,\cdots,u^{l(i,t)}]$, where $l(i,t)$ denotes the number of users the item $i$ was interacted before time $t$. 

Figure~\ref{fig:IPS_estimation} (a,b) illustrate that $\mathbf{h}^{<t}_u$ and $\mathbf{h}^{<t}_i$ are actually dual views of the data cube derived from $\mathcal{D}$. Specifically, by considering the user, item, and time as three axes, $\mathcal{D}$ can be represented as a sparse data cube where the $(u, i, t)$-th element is 1 if $u$ clicked $i$ at time $t$, 0 if observed but not clicked, and NULL if not-interacted.
It is obvious that $\mathbf{h}^{<t}_i$ and $\mathbf{h}^{<t}_u$ are two views of the data cube: (1) view of item: given a user $u$, her/his historical interactions before $t$ are stored in the matrix sliced by $u$. Since $u$ can only interact with one item at a time, we can remove the non-clicked items, sort the remaining items according to the time, and achieve the list $\mathbf{h}^{<t}_u$, where $l(u, t)$ is the number of nonzero elements in the sliced matrix; (2) view of 
user: given an item $i$ and its interaction history, the matrix sliced by $i$ can also be aggregated into another list $\mathbf{h}^{<t}_i$.

The task of sequential recommendation becomes, based on the user-item interactions and users' feedback in $\mathcal{D}$, learning a function $\hat{r}_t = f(u,i, \mathbf{h}_u^{<t},\mathbf{h}_i^{<t})$ that predicts user $u$'s preference on item $i$ at time $t$. It is expected that the predicted preference is close the true while un-observable user preference $r_t\in \{0,1\}$ at time $t$, where $r_t = 1$ means that $i$
is preferred by $u$ at time $t$, and 0 otherwise.

\subsection{Biases in Sequential Recommendation}
\begin{figure}
    \centering
    \includegraphics[width = 0.45\textwidth]{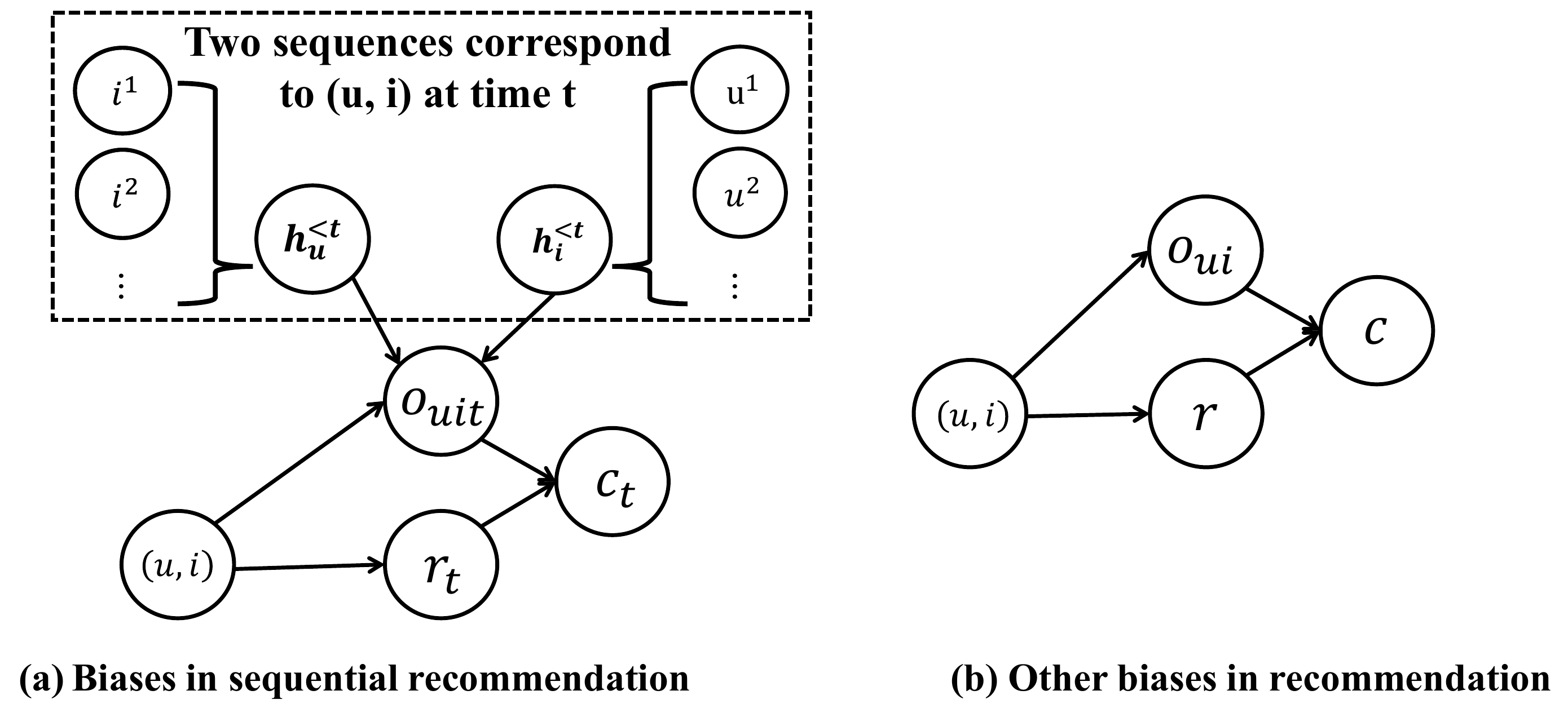}
    \caption{\mbox{Causal graphs of bias in (sequential) recommendation.}}
    \label{fig:CausalGraph}
\end{figure}
In sequential recommendation, bias happens when the user $u$ is systematically under/over-exposed to certain items. As shown in Figure~\ref{fig:CausalGraph}(a) and from a causal sense, a user clicks an item at time $t$ ($c_t = 1$) only if the item $i$ is relevant to the user $u$ ($r_t = 1$) and the $i$ is exposed to $u$ ($o_{uit} = 1$), where $o_{uit}\in\{0, 1\}$ and $r_t\in \{0,1\}$ respectively denote whether the user $u$ is aware of item $i$ and is relevant to $i$, or formally $c_t = r_t \cdot o_{uit}$. Further suppose that the two interaction sequences $\mathbf{h}^{<t}_u$ and $\mathbf{h}^{<t}_i$ %(respectively denoting $u$'s historical behaviors before $t$ and the item interacted with $i$ before $t$)
will also influence whether $u$ is aware of the $i$. Since the model predicts the user preference $r_t$ with the observed clicks $c_t$, the prediction is inevitably biased by the item exposure $o_{uit}$. This is because the 
$o_{uit}$ becomes a confounder after observing the click in the causal graph (i.e., click as a collider~\cite{pearl:09causal}). 

Formally, the probability of a user $u$ clicks an item $i$ at time $t$ can be factorized as the probability that $i$ is observable to the user (i.e., $o_{uit}=1$) and the probability that $(u,i)$ is relevant $r_t=1$\footnote{We suppose that the observational probability $P(o = 1)$ is only based on the user and item's historical interaction sequences. In real tasks, item observable probability is also influenced by other factors such as the ranking position, as shown in Figure~\ref{fig:CausalGraph}(b). Considering these factors in sequential recommendation will be the future work.}:
\begin{equation}\label{eq:click}
    P(c_t = 1|u,i,t ) = 
    P(r_t = 1|u,i,\mathbf{h}^{<t}_u,\mathbf{h}^{<t}_i)\cdot P(o_{uit} = 1).
\end{equation}
%by formulating the causal graph into a probability event, we write the tasks with the Equation~\ref{eq:click}, where $p(\cdot)$ denote the probability of the event happening.
% \xujun{
% Intuitively, a user would be hard to be exposed with his unaware items $P(o_{uit} = 1|\mathbf{h}^{<t}_u,\mathbf{h}^{<t}_i)$ caused by her/his previous behaviors $\mathbf{h}^{<t}_u$ and the communities that often interacted with $i$ $\mathbf{h}^{<t}_i$. Ideally, to help the users, the recommendation models should take both $\mathbf{h}^{<t}_u$ and $\mathbf{h}^{<t}_i$ into consideration. Existing de-biasing methods, however, are developed under the non-sequential recommendation scenarios and can only consider one of these two factors. }
%In the previous researches, the key to estimate the observation probability 
As have shown in Figure~\ref{intro}, existing studies usually estimate $P(o_{uit}=1)$ only from the view of item, ignoring the dual view of user. Also, the estimation need to take the chronological information (i.e., $\mathbf{h}^{<t}_u$ and $\mathbf{h}^{<t}_i$) into consideration.

\subsection{Unbiased Objective for Recommendation} %Propensity Score for Non-sequential Recommendation Scenarios}
%Unbiased recommendation has been proposed to address vairant biases in recommender systems. 
Ideally, the learning objective for recommendation (including the sequential and non-sequential scenarios) should be constructed based on the correlations between the true preference and the predicted score by the recommendation model:
\begin{equation}\label{eq:Non-sequential-objectivefunction}
    \mathcal{L}^{\text{ideal}}_{n} = \sum_{u\in \mathcal{U}}\sum_{i\in \mathcal{I}}\delta(r,\hat{r}(u,i)),
\end{equation}
where $\hat{r}(u,i)$ is the prediction by the recommendation model, $r$ is the true preference, and $\delta(\cdot,\cdot)$ is the loss function defined over each user-item pair. Note that in non-sequential recommendation, time $t$ and historical information $\mathbf{h}_u^{<t},\mathbf{h}_i^{<t}$ are not considered. In real world, however, $ \mathcal{L}^{\text{ideal}}_n$ cannot be directly optimized because the true preference $r$ cannot be observed.  

Traditional RS regard the observed clicks $c$ as the labels to learn the models, which is inevitably influenced by the exposure or self-selection bias as shown in Figure~\ref{fig:CausalGraph}(b).
A practical solution is to remedy the biases through the propensity score of the confounder~\cite{Survy:Unbias-Rec}. A typical approach developed under the non-sequential scenarios is utilizing the propensity score $P(o=1)$ to weigh each observed interaction:
\begin{equation}\label{eq:Non-sequential-Unbias_loss}
    \mathcal{L}^{\text{unbiased}}_n = \sum_{u\in \mathcal{U}}\sum_{i\in \mathcal{I}}\left[\mathbb{I}(o_{ui}=1)\frac{\delta(c,\hat{r}(u,i))}{P(o_{ui}=1)}\right],
\end{equation}
where $\mathbb{I}(\cdot)$ is the indicator function,
and 
Eq.~(\ref{eq:Non-sequential-Unbias_loss}) is an unbiased estimation of the ideal objective Eq.~(\ref{eq:Non-sequential-objectivefunction}), i.e., $\mathbb{E}_o\left[\mathcal{L}^{\mathrm{unbiased}}_n\right]$ = $\mathcal{L}^{\mathrm{ideal}}_n$. Please refer to~\cite{Survy:Unbias-Rec} for more details. 

Unbiased recommendation models have been developed under the framework. Generalizing these methods to sequential recommendation is a non-trial task. In this paper, we presented an approach to utilizing the user-item interaction sequences to estimate the propensity scores framework called DEPS.
% \begin{theorem}\label{theo:objective}
% $\mathcal{L}^{\mathrm{unbiased}}$ = $\mathcal{L}^{\mathrm{ideal}}$.
% \end{theorem}

\section{Our Approach: DEPS }
%In sequential recommendation, however, the learning objective is different and the propensity score $P(o=1)$ is hard to estimate. 
In this section, we proposed an unbiased objective for the sequential recommendation. After that, a dually enhanced IPS estimation model called DEPS is developed to estimate the propensity scores in the objective of sequential recommendations. Finally, two transformers are proposed to adapt our framework in a practical way. 

% Eq.~(\ref{eq:Unbias_loss}) cannot be directly optimized because the propensity score $P(o=1)$ is unobservable in recommendation. One solution is estimating the propensity score based on the users' past activities. In this section, we propose to estimate $P(o=1)$ based on both of the two aforementioned sequences: $\mathbf{h}^{<t}_i$ and $\mathbf{h}^{<t}_u$. A time-aware de-baising model called DEPS is developed which makes use of a two-sided transformer to estimate the probability. %in an unbiased and low variance way.    

\subsection{\mbox{Unbiased Loss for Sequential Recommendation}}
Given a user-item historical interactions $\mathcal{D}$, we define the ideal learning objective of sequential recommendation as evaluating the preference at each time that the users access the system:

\begin{equation}\label{eq:item-ideal-loss}
    \mathcal{L}^{\text{ideal}}_s = \sum_{u\in \mathcal{U}}\sum_{t:(u,i',c_t)\in\mathcal{D}}\sum_{i\in\mathcal{I}}\delta(r_t,\hat{r}_t(u,i,\mathbf{h}^{<t}_i,\mathbf{h}^{<t}_u)),
\end{equation}
where $\hat{r}(u,i, \mathbf{h}_i^{<t}, \mathbf{h}_u^{<t})$ is the prediction by the sequential recommendation model, and $r_t$ is the true while un-observable preference at time $t$. 
%where $(u,t)\sim \mathcal{D}$ means that according the records in $\mathcal{D}$, the user $u$ interacted with some item at time $t$.  
Different from the non-sequential unbiased recommender models, the propensity score in sequential recommendation is related to the time, as shown in the causal graph in Figure~\ref{fig:CausalGraph}(a). One way to achieve the unbiased sequential recommendation learning objective is estimating the propensity score $P(o_{uit}=1)$ corresponds to $(u,i)$ at time $t$, based on the historical interaction sequences of $\mathbf{h}^{<t}_i$ and $\mathbf{h}^{<t}_u$, as shown in the following theorem. 

%Assuming that an item $i$ can be observed ($o=1$) only if it appeared in the historical user sequence $\mathbf{h}_u^{<t}$, that is, $P(o=1) = P(i|\mathbf{h}_u^{<t})$. From the   
% can be estimated in a sequential manner, based on , we show that it can be estimated either from the users' view or the items' view of the user-system historical interactions, achieving two versions of the time-aware propensity score estimations, and is shown in the following Theorem~\ref{theo:ProbEstiamtion}:
%If the propensity score $P(o=1)$ only relates with the sequential information, that is,  the item $i$ will be observed (i.e. $o=1$) only if it appeared in the historical user sequence $h_u^{<t}$. And in dual property, the $o=1$ also implies the user $u$ appeared in the item historical sequence $h_i^{<t}$.

\begin{theorem}[Time-aware Unbiased Learning Objective]\label{theo:ProbEstiamtion}
Given user-item interactions $\mathcal{D}=\{(u,i, c_t)\}$, we have  %we have the $\mathcal{L}^{\mathrm{unbiased}}_s$ as an unbiased estimator for $\mathcal{L}^{\text{ideal}}_s$, that is,
\begin{equation}\label{eq:UnbiasLI_LU}
\mathbb{E}_o\left[\mathcal{L}^{\mathrm{unbiased}}_s\right] =\mathbb{E}_o\left[\alpha\mathcal{L}_u +(1-\alpha)\mathcal{L}_i\right] = \mathcal{L}^{\mathrm{ideal}}_s,
\end{equation}
where $\alpha\in[0,1]$ is the co-efficient that balances the two objectives:
\[
\mathcal{L}_u = \sum_{(u,i,c_t)\in\mathcal{D}}\left[\frac{\delta(c_t,\hat{r}_t)}{P(i,\mathbf{h}^{<t}_u)}\right] = \sum_{u\in\mathcal{U}}\sum_{(i,c_t)\in\mathcal{D}^u}\frac{\delta(c_t,\hat{r}_t)}{P(i,\mathbf{h}^{<t}_u)},
\]
\[
\mathcal{L}_i = \sum_{(u,i,c_t)\in\mathcal{D}}
\left[\frac{\delta(c_t,\hat{r}_t)}{P(u,\mathbf{h}^{<t}_i)}\right] = \sum_{i\in\mathcal{I}}\sum_{(u,c_t)\in\mathcal{D}^i}\frac{\delta(c_t,\hat{r}_t)}{P(u,\mathbf{h}^{<t}_i)},
%\mathcal{L}_u = \sum_{i\in\mathcal{I}}\sum_{u,t:(u,i,t)\sim\mathcal{D}}\left[\frac{\delta(c,\hat{r}(u,i,t))}{P(u|\mathbf{h}^{<t}_i)}\right].
\]

where $\mathcal{D}^u=\{(i,c_t):(u,i,c_t)\in\mathcal{D}\}$, $\mathcal{D}^i=\{(u,c_t):(u,i,c_t)\in\mathcal{D}\}$, %means that in the records in $\mathcal{D}$, $u$ interacted with $i$ at time $t$, 
$P(i,\mathbf{h}^{<t}_u)$ is the probability that $i$ and $\mathbf{h}^{<t}_i$ appear, and $P(u,\mathbf{h}^{<t}_i)$ is the probability that $u$ and $\mathbf{h}^{<t}_i$ appear.%\xujun{我的理解是这个i不一定要在h中出现}
\end{theorem}
Proof of Theorem~\ref{theo:ProbEstiamtion} can be found in the Appendix~\ref{sec:ProofTheorem1}. %The proof of Theorem~\ref{theo:ProbEstiamtion} also implies that under the sequential recommendation scenario, the propensity score $P(o=1)$ can be estimated as $P(i|\mathbf{h}^{<t}_u)$ or $P(u|\mathbf{h}^{<t}_i)$.
% \[
% P(o=1) = \frac{P(i|\mathbf{h}^{<t}_u) + P(u|\mathbf{h}^{<t}_i)}{2}. 
% \]
From the theorem, we can see that an unbiased learning objective for sequential recommendation can be achieved either from the view of user $\mathcal{L}_u$ or the view of item $\mathcal{L}_i$. 
Moreover, it is easy to know that the average of the two unbiased losses, i.e., $\mathcal{L}_s^\text{unbiased}$ defined in Eq.~(\ref{eq:UnbiasLI_LU}), is still an unbiased objective. 

Considering that the propensity scores play as the denominators in $\mathcal{L}_i$ and $\mathcal{L}_u$. To enhance the estimation stability, clip technique~\cite{UIR} is applied to the estimated probabilities in Eq.~(\ref{eq:prob_item}) and Eq.~(\ref{eq:prob_user}), achieving
\begin{align}
\label{eq:clip_prob_item} 
\widetilde{P}(i,\mathbf{h}^{<t}_u) =&\max\{P(i,\mathbf{h}^{<t}_u),M\},\\
\label{eq:clip_prob_user} 
\widetilde{P}(u,\mathbf{h}^{<t}_i) =&\max\{P(u,\mathbf{h}^{<t}_i),M\},
\end{align}
where $M\in (0, 1)$ is the clip value. We show that with the clipped propensity scores the estimation variance can be bounded:
\begin{theorem}[Estimation Variance]\label{theo:var}
Let $L_{u}^t = \frac{\delta(c_t,\hat{r}_t)}{\widetilde{P}(u,\mathbf{h}^{<t}_i)}$ and $L_{i}^t = \frac{\delta(c_t,\hat{r}_t)}{\widetilde{P}(i,\mathbf{h}^{<t}_u)}$ are two random variables w.r.t. loss on a single training sample $(i, u, c_t)\in\mathcal{D}$, $\alpha L_{u}^t+(1-\alpha)L_{i}^t$'s estimation variance satisfies:
\[
    \mathbb{V}\left[\alpha L_{u}^t+(1-\alpha)L_{i}^t\right] \leq  \max\{\mathbb{V}\left[L_{u}^t\right],\mathbb{V}\left[L_{i}^t\right]\} \leq \left(\frac{1}{M}-1\right)\delta^2(r_t,\hat{r}_t). 
\]
%and it is easy to prove that combining two views would not bring external variance, that is:
%\[
%    \mathbb{V}\left[\frac{L_{u}^t+L_{i}^t}{2}\right] \leq \max\{L_{u}^t,L_{i}^t\}.
%\]
\end{theorem}
Proof of Theorem~\ref{theo:var} can be found in the Appendix~\ref{sec:ProofTheorem2}. We conclude that the averaged loss would not bring additional variance. Intuitively, the clip value $M$ is a trade-off between the unbiasedness and the variance and provides a mechanism to control the variance. A larger $M$ leads to lower variance and more bias. We show that the clip technique in ~\citet{UIR} still works in a dual perspective.
%Similar results have also been reported in~\cite{UIR}. 

%Moreover, according to Cauchy-Schwarz' inequality, it is easy to know that the averaged loss would not bring additional variance because
% \[
%     \mathbb{V}\left[\frac{(L_{i}^t+L_{u}^t)}{2}\right]
%     \leq \frac{\mathbb{V}\left[L_{i}^t\right]+\mathbb{V}\left[L_{u}^t\right]+2\sqrt{\mathbb{V}\left[L_{i}^t\right]\mathbb{V}\left[L_{u}^t\right]}}{4}\leq \max\{L_{u}^t,L_{i}^t\}.
% \]
\begin{figure}
    \centering
    \includegraphics[width=0.45\textwidth]{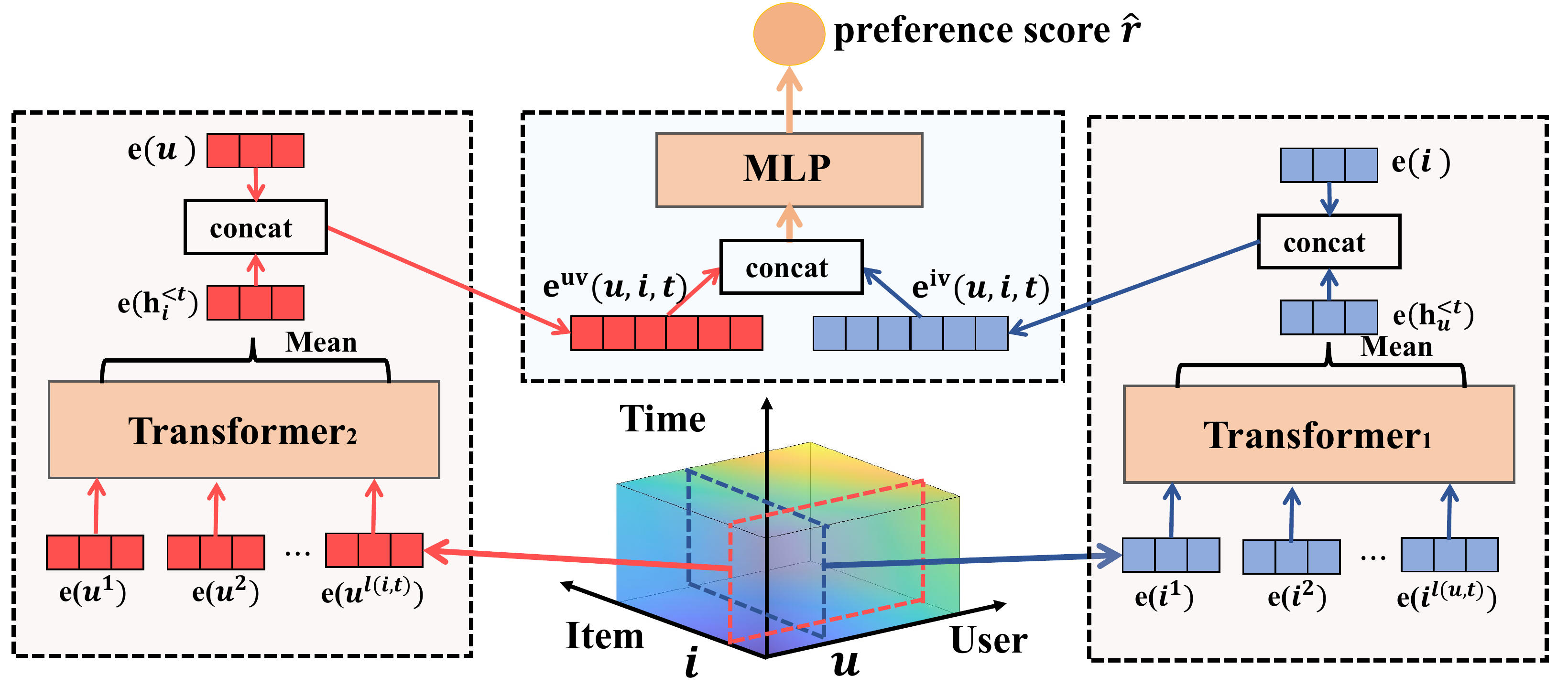}
    \caption{Two transformers based sequential recommender model with sequences from both users and items.}
    \label{fig:framework}
%    \vspace{-0.5cm}
\end{figure}

The above analysis provides an elegant and theoretically sound approach to learning an unbiased sequential recommendation model with three steps: (1) for each tuple $(u, i, \mathbf{h}^{<t}_u,\mathbf{h}^{<t}_i)$, estimating two propensity scores $\widetilde{P}(i,\mathbf{h}^{<t}_u)$ and $\widetilde{P}(u,\mathbf{h}^{<t}_i)$; (2) developing a sequential recommender model, and (3) training the parameters which involves minimizing the averaged loss $\alpha\mathcal{L}_i+(1-\alpha)\mathcal{L}_u$.
Next, we show an implementation of the step (1) with two GRUs in Section~\ref{sec:GRU4IPS}. Secton~\ref{sec:SR_DEPS} and Section~\ref{sec:TrainAlgorithm} respectively implement the step (2) and step (3). 

%\begin{theorem}\label{theo:IPS4SRS}
%The propensity score estimations $p(u|\mathbf{h}^{<t}_i)$ and $p(i|\mathbf{h}^{<t}_u)$ both are the unbiased estimation of propensity score, which also denotes that $\mathcal{L}^{unbiased}  = (\mathcal{L}^{unbiased}_i+\mathcal{L}^{unbiased}_u)/2$.
%\end{theorem}

%$p(i|\mathbf{h}_u^{<t}),p(u|\mathbf{h}_i^{<t})$ are both unbiased estimation of propensity score.

%One major challenge of constructing $\mathcal{L}^{\mathrm{unbiased}}$ is the unbiased estimation of the propensity score $P(o=1)$ under the sequential scenario. Theorem~\ref{theo:IPS4SRS} shows that $p(i|\mathbf{h}_u^{<t}),p(u|\mathbf{h}_i^{<t})$ are both unbiased estimation of propensity score. Moreover, theorem~\ref{theo:var} shows dual methods can control the  estimation variance well. On this theoretical analysis, a sequence predictor model is used to estimate the propensity score in an end-to-end style. 

%\subsubsection{Theoretical Analysis}

\subsection{Estimating Propensity Scores with GRUs}\label{sec:GRU4IPS}
%Theoretically, $P(i,\mathbf{h}^{<t}_u)$ and $P(u,\mathbf{h}^{<t}_i)$ can be calculated by simply counting the times that $i$ appears with sequence $\mathbf{h}^{<t}_u$ and the times that $u$ appears with sequence $\mathbf{h}^{<t}_i$. %, and then respectively divided by the corresponding sequence lengths. 
In the real world, sequences $\mathbf{h}^{<t}_u$ and $\mathbf{h}^{<t}_i$ are very sparse and short compared to the whole sets of items and users. %, resulting in outputting zero probabilities. 
In this paper, we resort to the neural language model of GRU~\cite{GRU} for estimating the propensity scores. 
Also, according to Theorem~\ref{theo:var}, the clip technique is applied to the estimated propensity scores.

Specifically, two GRUs are respectively used to estimate the $P(i,\mathbf{h}^{<t}_u)$ and $P(u,\mathbf{h}^{<t}_i)$, as shown in Figure~\ref{fig:IPS_estimation}(c). 
Specifically, given a tuple $(u, i, \mathbf{h}^{<t}_u, \mathbf{h}^{<t}_i)$, its propensity score from the view of item is estimated as the maximum value of $M$ and $P(i,\mathbf{h}^{<t}_u)$, where $P(i,\mathbf{h}^{<t}_u)$ is proportional to the dot product of $i$'s embedding $\mathbf{e}(i)$, and the output of a GRU which takes the sequence $\mathbf{h}^{<t}_u=[i^1, i^2, \cdots, i^{l(u,t)}]$ as input. By applying the clip technique in Eq.~(\ref{eq:clip_prob_item}), we write the estimated propensity score from view of item as:
\begin{equation}\label{eq:prob_item}
    % \widetilde{P}(i,\mathbf{h}^{<t}_u) =  \max\left\{ \frac{\mathbf{e}(i)^T\mathbf{y}(i^{l(u,t)})-\min_{{i'\in\mathcal{I}}}\mathbf{e}(i')^T \mathbf{y}(i^{l(u,t)})}{\max_{{i'\in\mathcal{I}}}\mathbf{e}(i')^T \mathbf{y}(i^{l(u,t)})-\min_{{i'\in\mathcal{I}}}\mathbf{e}(i')^T \mathbf{y}(i^{l(u,t)})},M\right\},
    \widetilde{P}(i,\mathbf{h}^{<t}_u) =  \max\left\{ \frac{\exp{\left(\mathbf{e}(i)^T\mathbf{y}(i^{l(u,t)})\right)}}{\sum_{i'\in\mathcal{I}}\exp{\left(\mathbf{e}(i')^T \mathbf{y}(i^{l(u,t)})\right)}},M\right\},
\end{equation}
where $\mathbf{y}(i^{l(u,t)})\in \mathbb{R}^d$ is the GRU output from its last layer (i.e., corresponds to the $l(u,t)$-th input). The GRU scans the items in $\mathbf{h}^{<t}_u$ as follows: at the $k$-th ($k=1, \cdots, l(u,t)$) layer, it takes the embedding of the $k$-th item $\mathbf{e}(i^{k})$ as input, and outputs $\mathbf{y}(i^{k})$ which is the representation for the scanned sub-sequence $[i^1, i^2, \cdots, i^k]$:
\[
    \mathbf{y}(i^{k}),\mathbf{z}^{k} = \textbf{GRU}_1(\mathbf{e}(i^{k}),\mathbf{z}^{k-1}),
\]
where $\mathbf{z}^k$ and $\mathbf{z}^{k-1}$ are the hidden vectors of $k$-th and $(k-1)$-th steps, and \textbf{GRU}$_1$ is the GRU cell that processes the sequence from the view of item.

Similarly, given a tuple $(u, i, \mathbf{h}^{<t}_u, \mathbf{h}^{<t}_i)$, its propensity score can also be estimated from the view of user: the maximum of $P(u,\mathbf{h}^{<t}_i)$ and $M$, where $P(u,\mathbf{h}^{<t}_i)$ is proportional to the dot product of the user embedding $\mathbf{e}(u)$ and the representation of sequence $\mathbf{h}^{<t}_i=[u^1, u^2, \cdots, u^{l(i,t)}]$. By applying the clip technique in Eq.~(\ref{eq:clip_prob_user}), we write the estimated propensity score from view of user as:
\begin{equation}\label{eq:prob_user}
    \widetilde{P}(u,\mathbf{h}^{<t}_i) =\max\left\{ \frac{\exp{\left(\mathbf{e}(u)^T\mathbf{y}(u^{l(i,t)})\right)}}{\sum_{u'\in\mathcal{U}}\exp{\left(\mathbf{e}(u')^T\mathbf{y}(u^{l(i,t)})\right)}},M\right\},
\end{equation}
where $\mathbf{y}(i^{l(u,t)})\in\mathbb{R}^d$ is the output of another GRU which scans $\mathbf{h}^{<t}_i$ as follows: at the $k$-th layer, it takes the embedding of the $k$-th user $\mathbf{e}(u^k)$ as input, and output $ \mathbf{y}(u^{k})$ representation for the scanned sub-sequence $[u^1, u^2, \cdots, u^k]$: 

\[
    \mathbf{y}(u^{k}),\mathbf{z}^{k} = \textbf{GRU}_2(\mathbf{e}(u^{k}),\mathbf{z}^{k-1}),
\]
where $\mathbf{z}^k$ and $\mathbf{z}^{k-1}$ are the hidden vectors, and $\textbf{GRU}_2$ is another GRU that processes the interaction sequence from the view of user. 

%, where $\theta_u,\theta_i$ denotes the parameters of the two GRU

%Also note that though we present a GRU-based implementation for propensity scores, any neural language models (e.g., the attention-based models, etc.) can also be used here.

\subsection{\mbox{Backbone: Transformer-based Recommender}}\label{sec:SR_DEPS}
%In this section, we present an implementation of the sequential recommendation model that makes use of the two types of user-item interaction sequences. 
As shown Figure~\ref{fig:framework}, %the model takes a tuple $(u, i, t)$ as the input and outputs the prediction of the $u$'s preference on $i$ at time $t$. 
%to use both the view of user and view of item of the interaction data, 
the implementation of the sequential recommendation model consists of a Transformer Layer and a Prediction Layer. The Transformer Layer consists of two transformers~\cite{attention}. One converts the sequence $\mathbf{h}_u^{<t}$ and the target item $i$ into representation vector, and another converts the sequence $\mathbf{h}_i^{<t}$ and the target user $u$ into another representation vector. The Prediction Layer concatenates the vectors and makes the prediction with an MLP.

\subsubsection{Transformer Layer}
%As for the down-stream recommendation model, we apply the transformer architecture to aggregate recommendation information due to the successful implement of DIN~\cite{DIN} and BST~\cite{BST}.

The overall item representation of the input tuple $(u, i, t)$ can be represented as the concat of item id embedding and user historical sequence embeddings:
\begin{equation}
    \mathbf{e}^{\textrm{iv}}(u,i,t) = \left[\mathbf{e}(\mathbf{h}^{<t}_u)\|\mathbf{e}(i)\right],
\end{equation}
where operator `$\|$' concatenates two vectors, $\mathbf{e}(i)$ is the embedding of the items, $\mathbf{h}^{<t}_u=[i^1, i^2, \cdots, i^{l(u,t)}]$ is the user sequence related to the target user $u$, and $\mathbf{e}(\mathbf{h}^{<t}_u)$ is the vector that encodes the sequence, defined as the mean of the vectors outputted by a transformer:
\begin{equation}\label{eq:EmbedSeqI}
    \mathbf{e}(\mathbf{h}^{<t}_u) = \text{Mean}\left(\textbf{Transformer$_1$}\left([\mathbf{e}(i^1),\cdots,\mathbf{e}(i^{l(u,t)})]\right)\right),
\end{equation}
where `Mean' is the mean pooling operation for all the input vectors, and \textbf{Transformer}$_1$($\cdot$) is a transformer~\cite{attention} architecture.

%we get the encoded user vector $\mathbf{v}_u(u,i,t)$ through a user-oriented transformers:

%where `$\|$' denotes the concat operation for two vectors. For the $\mathbf{v}(\mathbf{h}^{<t}_i$ is obtained by mean pooling the encoded user sequences output from the transformers:

Similarly, the overall item representation of the input tuple $(u,i,t)$ can be represented as the concat of user id embedding and item historical sequence embeddings:
\begin{equation}
    \mathbf{e}^{\textrm{uv}}(u,i,t) = \left[\mathbf{e}(\mathbf{h}^{<t}_i)\|\mathbf{e}(u)\right],
\end{equation}
$\mathbf{e}(u)$ is the embedding of the target user $i$, $\mathbf{h}^{<t}_i=[u^1, u^2, \cdots, u^{l(i,t)}]$ is item sequence interacted by the target item $i$, and $\mathbf{e}(\mathbf{h}^{<t}_i)$ is the mean of the output of another transformer:
\begin{equation}\label{eq:EmbedSeqU}
   \mathbf{e}(\mathbf{h}^{<t}_i) = \text{Mean}\left(\textbf{Transformer$_2$}\left([\mathbf{e}(u^1),\cdots,\mathbf{e}(u^{l(i,t)})]\right)\right),
\end{equation}
where \textbf{Transformer}$_2$($\cdot$) is another transformer architecture.

\subsubsection{Prediction Layer}
Finally, an Multi-Layer Perception (MLP) is applied which takes $\mathbf{e}^{\textrm{iv}}(u,i,t)$ and $\mathbf{e}^{\textrm{uv}}(u,i,t)$ as inputs, and outputs the predicted preference $\hat{r}$:
\begin{equation}\label{eq:prediction}
   \hat{r} = \sigma(\textbf{MLP}(\mathbf{e}^\text{iv}(u,i,t)|| \mathbf{e}^{\text{uv}}(u,i,t))),
\end{equation}
where `$\sigma$' is the sigmoid function operation and \textbf{MLP} is a two-layer fully connected neural network that takes both $\mathbf{e}^\text{iv}(u,i,t)$ and  $\mathbf{e}^{\text{uv}}(u,i,t)$ as inputs.

\begin{algorithm}[t]
    \caption{Learning Algorithm of DEPS}
	\label{alg:DEPS}
	\begin{algorithmic}[1]
	\REQUIRE Training set $\mathcal{D} = \{(u,i,c_t)\}$, iteration numbers $n_p, n_u, n_b$, coefficients $\lambda_p$
	
	\ENSURE $\mathbf{\Theta} = \{\theta_e,\theta_p,\theta_t,\theta_m\}$
	
	\STATE $\mathbf{\Theta}\leftarrow$ random values
	\STATE $\mathcal{H}\leftarrow \left\{\mathbf{h}_u^{<t}|u\in\mathcal{U}\right\}\cup \left\{\mathbf{h}_i^{<t}|i\in\mathcal{I}\right\}$\COMMENT{Extract seq. from $\mathcal{D}$}\\
%    \COMMENT{First stage learning}
	\FOR{$n=1,\cdots,n_p$}
%	    \STATE Extract the sequence $\mathbf{h}_u^{<t},\mathbf{h}_i^{<t}$ from $(u,i,t)$ 
	    \STATE  Update $\theta_p$ by minimizing $\mathcal{L}_u^{\text{AR}}+\mathcal{L}_i^{\text{AR}}$ 
	    \STATE Update $\theta_e,\theta_t$ by minimizing $\lambda_p(\mathcal{L}^{{\text{MLM}}}_u +  \mathcal{L}^{{\text{MLM}}}_i)$
	\ENDFOR
	
%	\COMMENT{Second Stage Learning}
	\FOR{$n=1,\cdots,n_u$}
	    \STATE Extract the sequence $\mathbf{h}_u^{<t},\mathbf{h}_i^{<t}$ from $(u,i,t)$ 
	    \FOR{$k=1,\cdots,n_b$}
    	    \STATE  Update $\theta_p$ by minimizing the loss $\mathcal{L}_u^{{\text{AR}}}+\mathcal{L}_i^{{\text{AR}}}$.
    	\ENDFOR
	    \STATE Calculate preference score $\hat{r} = f(u,i,\mathbf{h}_u^{<t},\mathbf{h}_i^{<t})$ (Eq.~(\ref{eq:prediction})) 
	    \STATE Calculate propensity score $\widetilde{P}(i,\mathbf{h}^{<t}_u),\widetilde{P}(u,\mathbf{h}^{<t}_i)$ (Eq.~(\ref{eq:clip_prob_item}), (\ref{eq:clip_prob_user})) 
	    \STATE Update $\theta_e,\theta_t,\theta_m$ by minimizing the loss $\mathcal{L}^{\textrm{unbiased}}_s$.
	\ENDFOR
	\end{algorithmic}
	
\end{algorithm}

\subsection{Learning with Estimated Propensity Scores}\label{sec:TrainAlgorithm}
The proposed model has a set of parameters to learn, denoted as $\mathbf{\Theta} = \{\theta_e,\theta_p,\theta_t,\theta_m\}$, where parameters $\theta_e$ denotes the parameters in the embedding models which output the user and item embeddings, the parameters $\theta_p$ in \textbf{GRU}$_1$ and \textbf{GRU}$_2$ for estimating propensity scores, the parameters $\theta_t$ in \textbf{Transformer}$_1$, \textbf{Transformer}$_2$, and the parameters $\theta_m$ in \textbf{MLP} for making the final recommendation. 

Inspired by the pre-train and then fine-tune paradigm, we also design a two-stage learning procedure to learn the model parameters. In the first stage, the parameters of $\{\theta_e,\theta_p,\theta_t\}$ are trained in an unsupervised learning manner, achieving a relatively good initialization. Then, the second stage learns all of the parameters with the aforementioned unbiased learning objectives. Adam~\cite{adam} optimizer is used for conducting the optimization.

For stage-1, we apply $n_p$ epochs to optimize the $\mathcal{L}^\textrm{stage-1}$. For stage-2, we apply $n_u$ epochs to alternative train, where in each epoch, $n_b$ epochs to optimize the $\mathcal{L}^\textrm{AR}_u + \mathcal{L}^\textrm{AR}_i$ and 1 epoch is set to optimize the $\mathcal{L}^\textrm{unbiased}_s$, respectively. The overall algorithm process can be seen in Algorithm~\ref{alg:DEPS}.

%$n_b$ epochs to optimize the objective $\mathcal{L}^{AR}_u + \mathcal{L}^{AR}_i$ and apply

\subsubsection{First Stage: Unsupervised Learning}
In the first stage, the two views of all user-system interaction sequences, i.e., $\mathbf{h}^{<t}_i$'s and $\mathbf{h}^{<t}_u$'s for all $u\in\mathcal{U}$ and $i\in\mathcal{I}$, are utilized as the unsupervised training instances. Inspired by the success of the Autoregressive language models and the masked language models, two learning tasks are designed which respectively apply these two languages models to the user sequences and item sequences, resulting in a total loss $\mathcal{L}^\text{stage-1}$ that consists of four parts:
%in order to obtain a relatively good initialization of parameters and control the variance of the IPS estimation, we designed the unsupervised learning tasks based on the training data $\mathcal{D}$. Specifically, a total loss $\mathcal{L}^\text{stage-1}$ which consists of two types of learning tasks (in total four losses) is jointly optimized: 
\begin{equation}
    \mathcal{L}^\text{stage-1} = (\mathcal{L}^\text{AR}_u +\mathcal{L}^\text{AR}_i) + \lambda_p(\mathcal{L}^\text{MLM}_u +  \mathcal{L}^\text{MLM}_i),
\end{equation}
where $\lambda_p>0$ is the a trade-off coefficient, $\mathcal{L}^\text{AR}_u$ and $\mathcal{L}^\text{AR}_i$ are the losses correspond to respectively apply the Autoregressive language models to the sequences of $\mathcal{L}^\text{AR}_u$ and $\mathcal{L}^{AR}_i$. Specifically, $\mathcal{L}^\text{AR}_u$ is defined as:
\[
    \mathcal{L}^\text{AR}_u = \sum_{u\in \mathcal{U}}\text{AR}(\mathbf{h}^{<t}_u) = \sum_{u\in \mathcal{U}}\sum_{m=1}^{l(u,t)} -\log P\left(i^{m}|\left[i^1, \cdots, i^{m-1}\right]\right),
\]
where $i^m$ is the $m$-th item in sequence $\mathbf{h}^{<t}_u$ and $\left[i^1, \cdots, i^{m-1}\right]$ is the $(m-1)$-length prefix of $\mathbf{h}^{<t}_u$, and the probability $P(\cdot|\cdot)$ is calculated according to Eq.~(\ref{eq:prob_item}). 
Similarly, $\mathcal{L}^{AR}_i$ is defined as:
\[
    \mathcal{L}^\text{AR}_i = \sum_{i\in \mathcal{I}}\text{AR}(\mathbf{h}^{<t}_i) = \sum_{i\in \mathcal{I}}\sum_{m=1}^{l(i,t)} -\log P\left(u^{m}|\left[u^1, \cdots, u^{m-1}\right]\right),
\]
 probability $P(\cdot|\cdot)$ is calculated according to Eq.~(\ref{eq:prob_user}).

%maximize the estimated propensity scores outputted by Eq.~(\ref{eq:prob_item}), and $\mathcal{L}^\text{mask}_u$ and $\mathcal{L}^{mask}_i$ are designed to predict the masked users/items given others non-masked users or items. 

%Specifically, given a user $u$ and the interaction sequence derived from $\mathcal{D}$, the corresponding $\mathcal{L}^\text{IPS}_u$ is defined as the sum of the log probability of the each item at time step $l(i,t+1)$ given its former sequences $\mathbf{h}_i^{<t}$ \xujun{$i \in \mathbf{h}_u$?}
%\[
%    \mathcal{L}^{IPS}_u = \sum_{i\in \mathcal{I}}\sum_t-\log\left[P(u^{l(i,t+1)}|\mathbf{h}_i^{<t})\right]
%\]

%are designed to initialize the RNNs in Section~\ref{sec:GRU4IPS}, and 

%For the user IPS estimation optimize object $\mathcal{L}^{mask}_u, \mathcal{L}^{mask}_i$, we aims to train the parameters as a language model discussed in Section~\ref{sec:GRU4IPS}. Specifically, for a item sequences interacted with the user $u$,  we aims to maximize the log probability of the each item at time step $l(i,t+1)$ given its former sequences $\mathbf{h}_i^{<t}$, that is,
%\begin{equation}
%    \mathcal{L}^{IPS}_u = \sum_{i\in \mathcal{I}}\sum_t-log\left[p(u^{l(i,t+1)}|\mathbf{h}_i^{<t})\right]
%\end{equation}
%Similarly, for a user sequences interacting with the item $i$,  we aims to maximize the log probability of the each user at time step $l(u,t+1)$ given its former sequences $\mathbf{h}_u^{<t}$, that is,
%\begin{equation}
%    \mathcal{L}^{IPS}_i = \sum_{u\in \mathcal{U}}\sum_t-log\left[p(i^{l(u,t+1)}|\mathbf{h}_u^{<t})\right]
%\end{equation}.

As for $\mathcal{L}^\text{MLM}_u$ and  $\mathcal{L}^\text{MLM}_i$, following the practice in BERT4Rec~\cite{bert4rec}, we respectively apply the masked language models to the sequences of $\mathbf{h}^{<t}_u$ and $\mathbf{h}^{<t}_i$, achieving:
\[
\mathcal{L}^\text{MLM}_u = \sum_{u\in \mathcal{U}}\text{MLM}\left(\mathbf{h}^{<t}_u\right); \mathcal{L}^\text{MLM}_i= \sum_{i\in \mathcal{I}}\text{MLM}\left(\mathbf{h}^{<t}_i\right),
\]
where MLM($\cdot$) calculates the masked language model loss on the inputted sequence. The MLM task will make our training phase of the second stage more stable.
\subsubsection{Second Stage: Unbiased Learning}
In the second stage training, given $\mathcal{D} =\{(u, i, c_t)\}$, the estimated propensity scores is used to re-weight the original biased loss $\delta(c,\hat{r})$, achieving the unbiased loss $\mathcal{L}^\text{stage-2}$:
\begin{equation}\label{eq:stage2-loss}
    \mathcal{L}^\text{stage-2} = \mathcal{L}^\text{unbiased}_s,
\end{equation}
where unbiased objective $\mathcal{L}^\text{unbiased}_s$ is constructed based on the Eq.~(\ref{eq:UnbiasLI_LU}) in Theorem~\ref{theo:ProbEstiamtion}, by substituting the estimated propensity scores in Eq.~(\ref{eq:clip_prob_item}) and Eq.~(\ref{eq:clip_prob_user}) to Eq.~(\ref{eq:UnbiasLI_LU}). 

Note that the second-stage also empirically involves $\mathcal{L}^{AR}_u$ and $\mathcal{L}^{AR}_i$ for avoiding the high variance of propensity score estimation for alternate training. In all of the experiments of this paper, he original loss $\delta(c,\hat{r})$ was set to the binary cross entropy:
\[
    \delta(c,\hat{r}) = c\cdot \log(\hat{r}) + (1-c)\log(1-\hat{r}),
\]
where $\hat{r}$ is predicted by Eq.~(\ref{eq:prediction}).

\section{Experiments}
We conducted experiments to verify the effectiveness of DEPS.\footnote{The
source code is shared at~\url{https://github.com/XuChen0427/Dually-Enhanced-Propensity-Score-Estimation-in-Sequential-Recommendation}.}

% The experiments aim to answer the following research questions:

% \textbf{RQ1}: Can DEPS improve the accuracy for sequential recommendation? %  CORRESPONDING TO: 5.1

% % \textbf{RQ2}: How can DEPS improve recommendation?

% \textbf{RQ2}: How do the items and users in the interaction sequences help DEPS to break the information cocoons?  %CORRESPONDING TO: 5.3.1

% \textbf{RQ3}: Can the estimated propensity scores be used to improve other methods in a model-agnostic way? %CORRESPONDING TO: 5.3.2

\subsection{Experimental Settings}
The experiments were conducted on four large scale publicly available sequential recommendation benchmarks: 
 
\textbf{MIND}\footnote{\url{https://msnews.github.io/}}: a large scale news recommendation dataset. Users/items interacted with less than 5 items/users were removed for avoiding extremely sparse cases.

\textbf{Amazon-Beauty/Amazon-Digital-Music}: Two subsets (beauty and digital music domains) of Amazon Product dataset\footnote{\url{http://jmcauley.ucsd.edu/data/amazon/}}. Similarly, users/items interacted with less than 5 items/users were removed. We treated the 4-5 star ratings of Amazon dataset made by users as positive feedback (labeled with $1$), and others as negative feedback (labeled with $0$).

\textbf{Huawei Dataset}: To verify the effectiveness of our method on production data, we collect 1 month traffic log from the Huawei music service system, with about 245K interactions after sampling.

Table~\ref{tab:dataset} lists statistics of the four datasets. Following the practices in~\cite{bert4rec,FPMC}, Debiased recommender models need to be evaluated based on unbiased testing sets~\cite{UIR}. Following the practice of ~\cite{saito20:ATR}, we utilized the first 50\% interactions sorted by interaction times for training and re-sample other 50\% data for evaluation and test. Specifically, suppose item $i$ were clicked $m_i$ times, we used the inverse probability $m_i/\max_{j\in\mathcal{I}}m_j$ to sample. Then we utilized 20\% and 30\% sorted data for validation and test, respectively.

%For each user, the interactions by the user were grouped and sorted by the timestamp and split to ``train-valid-test'' by the leave-one-out splitting: the second last interaction for model validation and the last for the model test.

%we converted all numeric ratings to implicit feedback of 0/1 (e.g. rating $\ge 4$ to implicit feedback 1, otherwise 0). 
%The interaction by users were grouped and split to ``train-valid-test'' sets through the leave-one-out splitting: for each user, the last two interaction were set as a validation and a test record, respectively.
\begin{table}[t]
\small
\caption{Statistics of the datasets. } \label{tab:dataset}
\centering
\begin{tabular}{lcccc}
\hline
Dataset              & \#User & \#Item & \#Interaction & Sparsity \\
\hline
\hline
%\hline
MIND                 & 13863  & 2464  & 59228      & 99.82\% \\
%\hline
Amazon-Beauty        & 24411  & 32371  & 94641       & 99.98\% \\
%\hline
Amazon-Digital-Music & 4424   & 5365   & 32314        & 99.86\% \\
Huawei         & 1997   & 17490   & 245564      & 99.29\% \\
\hline
%\vspace{-0.8cm}
\end{tabular}
\end{table}

% \subsection{Baselines and Implementation Details}
% \label{sec:baseline}
The following representative sequential recommendation models were chosen as the {baselines}:
\textbf{STAMP}~\cite{stamp} which models the long- and short-term preference of users; \textbf{GRU4Rec+}~\cite{imporvedRec4gru} is an improved version of GRU4Rec with data augmentation and accounting for the shifts in the inputs; \textbf{BERT4Rec}~\cite{bert4rec} employs an attention module to model user behaviors and trains with unsupervised style;  \textbf{FPMC}~\cite{FPMC} captures users’ preference by combing matrix factorization with first-order Markov chains; \textbf{DIN}~\cite{DIN} applies an attention module to adaptively learn the user interests from their historical behaviors;  \textbf{BST}~\cite{BST} applies the transformer architecture to adaptively learn user interests from historical behaviors and the side information of users and items; \textbf{LightSANs}~\cite{lightsans} is a low-rank decomposed SANs-based recommender model. We also chose the following unbiased recommendation models as the baselines: \textbf{UIR}~\cite{UIR} is an unbiased recommendation model that estimates the propensity score using heuristics;  \textbf{CPR}~\cite{cpr} is a pairwise debiasing approach for exposure bias;  \textbf{UBPR}~\cite{ubpr}is an IPS method for non-negative pair-wise loss. \textbf{DICE}~\cite{dice}: A debiasing model focused on the user communities.
\textbf{USR}~\cite{wang2022unbiased}: A debiasing sequential model that aims to alleviate bias raised by latent confounders.

\begin{table*}[t]
\setlength{\tabcolsep}{4.5pt}
    \Small
        \caption{Performance comparisons between DEPS and the baselines on MIND, Beauty, Music, and Huawei datasets. `$*$' means the improvements over the best baseline (the underlined number) are statistical significant (t-tests and $p$-value $< 0.05$).}
    \label{tab:EXP:main}
    %\vspace{-0.2cm}
    \centering
    %\vspace{-0.2cm}
    \centering
    \begin{tabular}{|l@{}|l@{}|ccccccc|ccccc|cc@{}|}
        \hline
        \multicolumn{2}{|l|}{} & \multicolumn{7}{c|}{Sequential recommender baselines} & \multicolumn{5}{c|}{Unbiased recommender baselines} & \multicolumn{2}{c|}{Our approach} \\
        \hline
         \textbf{Dataset} & {Metric} & STAMP & DIN  & BERT4Rec & FPMC & GRU4Rec+ & BST & LightSANs & UIR & CPR & UBPR & DICE & USR & \textbf{DEPS} & Improv.\\
        \hline

\multirow{6}{*}{MIND}  & {NDCG@5} & 0.0471 & 0.1149  & 0.0900 & 0.0670 & 0.0865 & 0.0865 & \underline{0.1148} & 0.0594 & 0.0582 & 0.0588 & 0.0612 & 0.0658 &  \textbf{0.1197}* & 4.2\% \\
& {NDCG@10} & 0.0669 & 0.1548 & 0.1277 & 0.1006 & 0.1306 & 0.1233 & \underline{0.1650} & 0.0823 & 0.0847 & 0.0863 & 0.0861 & 0.0955 & \textbf{0.1728}* & 4.7\% \\
& {NDCG@20} & 0.0997 & 0.1948 & 0.1817 & 0.1400 & 0.1819 & 0.1753 & \underline{0.2159} & 0.1233 & 0.1204 & 0.1237 & 0.1235 & 0.1339 & \textbf{0.2249}* & 4.2\% \\
& {HR@5} & 0.0861 & 0.2090 & 0.1671 & 0.1229 & 0.1504 & 0.1607 & \underline{0.2024} & 0.1141 & 0.1037 & 0.1048 & 0.1201 & 0.1207 & \textbf{0.2200}* & 8.7\%\\
& {HR@10} & 0.1519 & 0.3379 & 0.2922 & 0.2359 & 0.2918 & 0.2825 & \underline{0.3692} & 0.1909 & 0.1898 & 0.1942 & 0.2030 & 0,2194 &\textbf{0.3961}* & 7.3\% \\
& {HR@20} & 0.2863 & 0.5019 & 0.5145 & 0.3999 & 0.5052 & 0.4948 & \underline{0.5760} & 0.3577 & 0.3379 & 0.3511 & 0.3571 & 0.3807 & \textbf{0.6078}* & 5.5\%  \\
\hline
  & {NDCG@5} & 0.0985 & 0.1139 & 0.1008 & 0.1225 & 0.1050 & 0.1156 & \underline{0.1312} & 0.1188 & 0.1172 & 0.0918 & 0.1238 & 0.1046 & \textbf{0.1362}* & 3.8\% \\
& {NDCG@10} & 0.1330 & 0.1407 & 0.1333 & 0.1563 & 0.1438 & 0.1548 & \underline{0.1661} & 0.1603 & 0.1465 & 0.1214 & 0.1645 & 0.1429 & \textbf{0.1830}* & 10.2\% \\
{Amazon-}& {NDCG@20} & 0.1780 & 0.1699 & 0.1722 & 0.1962 & 0.1919 & 0.1983 & 0.2078 & 0.2109 & 0.1823 & 0.1580 & \underline{0.2127} & 0.1903 & \textbf{0.2302}* & 8.2\% \\
{Beauty}& {HR@5} & 0.1945 & 0.2160 & 0.1964 & 0.2292 & 0.2132 & 0.2134 & 0.2430 & 0.2338 & 0.2119 & 0.1780 & \underline{0.2365} & 0.2064 & \textbf{0.2557}* &  8.1\% \\
& {HR@10} & 0.3349 & 0.3174 & 0.3166 & 0.3530 & 0.3555 & 0.3679 & 0.3696 & \underline{0.3836} & 0.3176 & 0.2879 & 0.3834 & 0.3435 & \textbf{0.4175}* & 8.8\% \\
& {HR@20} & 0.5364 & 0.4434 & 0.4864 & 0.5233 & 0.5656 & 0.5524 & 0.5424 & \underline{0.5980} & 0.4756 & 0.4465 & 0.5890 & 0.5470 & \textbf{0.6130}* & 2.5\% \\
\hline
  & {NDCG@5} & 0.0727 & 0.1220 & 0.1419 & 0.1355 & 0.0651 & 0.1556 & \underline{0.1940} & 0.0691 & 0.1664 & 0.1353 & 0.1170 & 0.0921 & \textbf{0.2256}* &  16.3\% \\
 {Amazon-}& {NDCG@10} & 0.0961 & 0.1513 & 0.1771 & 0.1799 & 0.0961 & 0.1976 & \underline{0.2192} & 0.0912 & 0.2121 & 0.1643 & 0.1432 & 0.1222 & \textbf{0.2766}* & 30.4\% \\
 {Digital-}& {NDCG@20} & 0.1305 & 0.1890 & 0.2231 & 0.2153 & 0.1385 & 0.2454 & \underline{0.2586} & 0.1317 & 0.2558 & 0.2060 & 0.1789 & 0.1643 & \textbf{0.3273}* & 26.6\%\\
 {Music}& {HR@5} & 0.1373 & 0.2745 & 0.2606 & 0.2941 & 0.1709 & 0.3249 & \underline{0.3613} & 0.1541 & 0.3473 & 0.2521 & 0.2185 & 0.2017 & \textbf{0.4093}* & 13.3\%\\
& {HR@10} & 0.2437 & 0.3950 & 0.4249 & 0.4391 & 0.3025 & 0.4902 & 0.4737 & 0.2437 &  \underline{0.5014} & 0.3754 & 0.3361 & 0.3305 & \textbf{0.5852}* & 16.7\% \\
& {HR@20} & 0.4034 & 0.5518 & 0.5949 & 0.6091 & 0.4874 & \underline{0.6779} & 0.6162 & 0.4314 & 0.6415 & 0.5546 & 0.4902 & 0.5098 & \textbf{0.7757}* & 14.4\% \\
\hline
\multirow{6}{*}{Huawei}  & {NDCG@5} & 0.0919 & 0.1081 & 0.1247 & 0.1079 & 0.1090 & 0.1016 & 0.1335 & 0.0554 & \underline{0.1414} & 0.1033 & 0.0670 & 0.0922 & 0.1400 & -1.0\%   \\
& {NDCG@10} & 0.1024 & 0.1195 & 0.1352 & 0.1193 & 0.1163 & 0.1125 & 0.1323 & 0.0634 & \underline{0.1489} & 0.1192 & 0.0767 & 0.1042 & \textbf{0.1503}* & 1.0\%   \\
& {NDCG@20} & 0.1202 & 0.1421 & 0.1561 & 0.1416 & 0.1357 & 0.1344 & 0.1454 & 0.0793 & \underline{0.1711} & 0.1409 & 0.0926 & 0.1234 & \textbf{0.1755}* & 2.5\%   \\
& {HR@5} & 0.3569 & 0.3874 & 0.4192 & 0.3954 & 0.3941 & 0.3594 & 0.4576 & 0.2129 & \underline{0.4726} & 0.3880 & 0.2453 & 0.3435 & \textbf{0.4765}* & 0.9\% \\
& {HR@10} & 0.5540 & 0.6138 & 0.6242 & 0.6174 & 0.5833 & 0.5625 & 0.6315 & 0.3569 & \underline{0.6730} & 0.6132 & 0.4155 & 0.5552 & \textbf{0.6852}* & 1.8\%\\
& {HR@20} & 0.7404 & 0.8011 & 0.8090 & 0.8188 & 0.7890 & 0.7785 & 0.7889 & 0.5625 & \underline{0.8603} & 0.8115 & 0.6107 & 0.7529 & \textbf{0.8713}* & 1.3\% \\
\hline
    \end{tabular}
    %\vspace{-0.3cm}
\end{table*}

To evaluate the performances of DEPS and baselines, we utilized two types of metrics: the accuracy of recommendation~\cite{FPMC,bert4rec} in terms of NDCG@K and HR@K. Following the practices in~\cite{bert4rec,DIN}, for representing the users and items in the sequences (i.e., users in $\mathbf{h}_i^{<t}$ and items in $\mathbf{h}_u^{<t}$), sequence position embedding were added to the original embedding. 

As for the hyper parameters in all models, the learning rate was tuned among $[1e-3,1e-4]$ and the propensity score estimation clip coefficient $M$ was tuned among $[0.01,0.2]$. The trade-off coefficients in the first-stage $\lambda_p$ was set to $0.5$. The trade-off coefficients $\alpha$ of two views was tuned among $[0.4,0.6]$.
The hidden dimensions of the neural networks $d$ was tuned among $\{64,128,256\}$, the dropout rate was tuned among $\{0.1,0.2,0.3,0.4,0.5\}$, and number of transformer layers was tuned among $\{2,3,4\}$. 

All the baselines and the experiments were developed and conducted under the Recbole recommender tools~\cite{recbole} and Pytorch~\cite{pytorch}. All the models were trained on a single NVIDIA GeForce RTX 3090, with the batch size tuned among $\{1024,2048,4096\}$.

\subsection{Experimental Results}

Table~\ref{tab:EXP:main} reports the experimental results of DEPS and the baselines on all of the four datasets, in terms of NDCG@K and HR@K which measure the recommendation accuracy. `$*$' means the improvements over the best baseline are statistical significant (t-tests and $p$-value $< 0.05$). Underlines indicate the best-performed methods. 

From the reported results, we can see that DEPS significantly outperformed nearly all of the baselines in terms of NDCG and HR expect NDCG@5 on Huawei commercial data, verified the effectiveness of DEPS in terms of improving the sequential recommendation accuracy. 
Moreover, DEPS significantly outperformed the unbiased models, demonstrating the importance of estimating propensity scores from the views of item and user \mbox{ sequential recommendation.}

% At the same time, we also adapted our framework DEPS to improve other sequential recommendation models. Please note that the sequential recommendation models of GRU4Rec and FPCM cannot be trained on MLM tasks. Therefore, in the first stage training of DEPS (GRU4Rec+) and DEPS (FPMC), the loss degenerates to $\mathcal{L}_u^{\text{AR}}+\mathcal{L}_i^{\text{AR}}$. From the NDCG@K and HR@K scores reported in Table~\ref{tab:EXP:main}, we found that the DEPS (GRU4Rec) and DEPS (FPMC) respectively achieved significant improvements over their underlying models of GRU4Rec+ and FPMC, indicating that the propensity scores estimated by DEPS are general. They can be used to improve other sequential recommendation models in a model-agnostic manner. 

% \begin{figure*}[ht]  
%     \centering    
%     \subfigure[Ablation study for dual IPS methods]
%     {
%         \includegraphics[width=0.45\linewidth]{image/ablation_ips.pdf}
%     }
%     \subfigure[Ablation study for dual transformers]
%     {
%         \includegraphics[width=0.45\linewidth]{image/ablation_trans.pdf}
%     }
%     \vspace{-0.3cm}
%     \caption{Empirical analysis based on Amazon-Digital Music. (a):  DEPS(trans.) variation performances of NDCG@K and HR@K (b): transfomer variation DEPS(trans.)s' performance of NDCG@K and HR@K}
%     \label{fig:ablation}  
% \end{figure*}

\begin{figure}[t]  
    \centering    
    \includegraphics[width=0.9\linewidth]{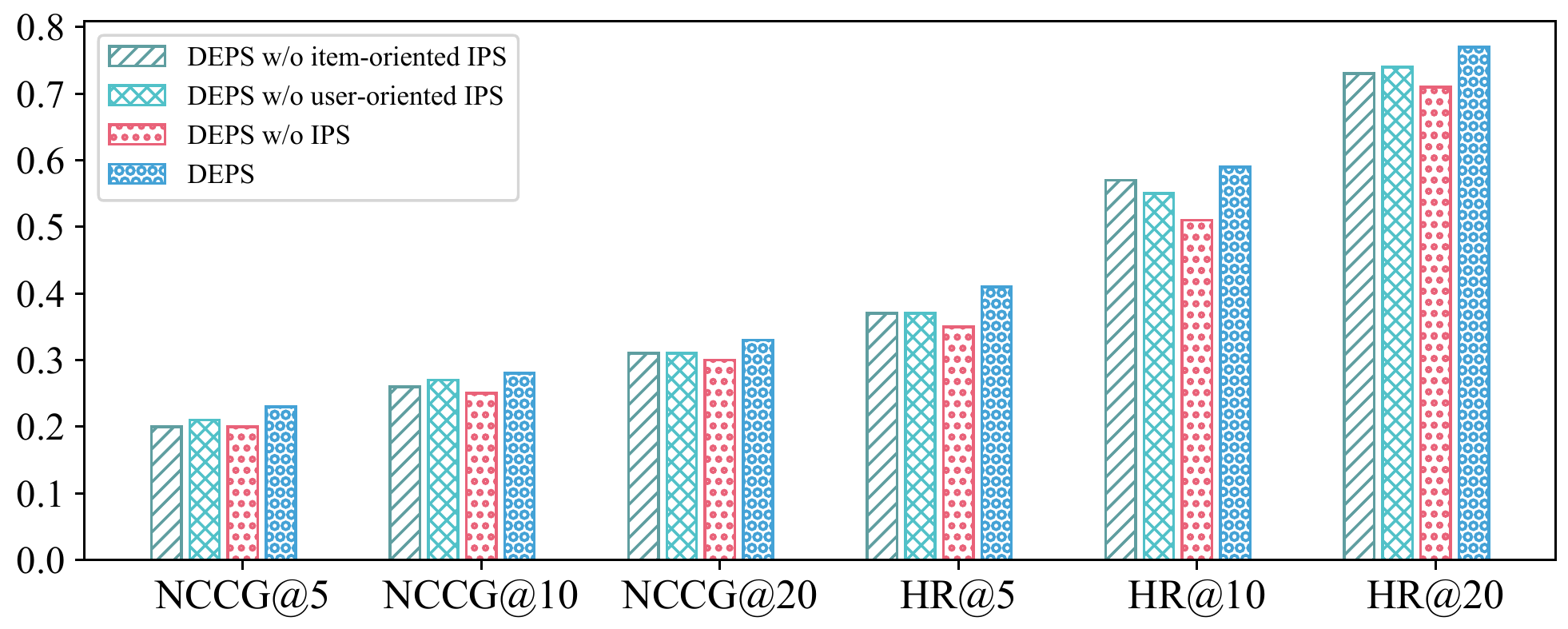}
    \caption{Empirical analysis based on Amazon-Digital Music:  DEPS variation performances of NDCG@K and HR@K}
    \label{fig:ablation_ips}  
\end{figure}

%From the diversity results, we can see that diversity of DEPS outperformed most of the baselines significantly. The results indicate that DEPS recommended more diverse items (lower GiniIndex@10 and higher Coverage@10) than other recommender models with a large margin. Therefore, the users were exposed to more diverse items so they would be less likely to be stuck in information cocoons. 

%\xujun{We need to show why DEPS can outperform these types of models. What are the conclusions? }

% \begin{table}[t]
%     \small
%     \caption{Item diversity experiment on MovieLen-1M dataset by uniformly sampling 100 items in item set. }   
%     \label{tab:diversity}
%     %\vspace{-0.2cm}
%     \centering
    
%     \begin{tabular}{l|c|c}
%         \hline
%          \textbf{Models}  & GiniIndex@10 (\%) & Coverage@10 (\%) \\
%         \hline \hline
%         DeepFM~\cite{deepfm} & 88.47 & 22.17 \\
%         GRU4Rec+~\cite{imporvedRec4gru} &  87.99 & 35.18\\
%          DMF~\cite{DMF} & 83.31 & 39.12 \\
%         NCF~\cite{NCF} & 83.94 & 43.11 \\
%         BST~\cite{BST} & 84.38 & 40.65\\
%         UIR~\cite{UIR} & 81.62 & 45.37 \\
%         FPMC~\cite{FPMC} & 86.01 & 61.69 \\
%         DIN~\cite{DIN} & 66.17 & 84.35\\
%         BERT4Rec~\cite{bert4rec} & 58.2 & 87.19 \\
%         \hline
%         \textbf{DEPS(ours)} & \textbf{58.45} & \textbf{88.57} \\
%         \hline
%     \end{tabular}
% \end{table}

% Please add the following required packages to your document preamble:
% \usepackage{multirow}
\subsection{Experimental Analysis}
We conducted more experiments to analyze DEPS, based on the Amazon-Digital-Music test data. %Similar phenomenon has also been observed on other datasets and backbones.

\subsubsection{Ablation Study}\label{sec:abla}

To further show the importance of estimating propensity scores with the two types of sequences from the view of user and the view of item, we also studied their unbiased performance in the second stage of training when the $\mathcal{L}^{unbiased}$ is optimized. Specifically, we showed the NDCG@K and HR@K of several DEPS variations. These variations include learning the recommendation model with no propensity score estimation (denoted as ``w/o IPS''), estimating propensity scores with view of item sequences only (``w/o user-oriented IPS''), with the view of user sequences only (``w/o item-oriented IPS''). 
From the performances shown in Figure~\ref{fig:ablation_ips}, we found that (1) ``w/o propensity'' performed the worst, indicating the importance of propensity scores are in unbiased sequence recommendation; (2) ``w/o user-oriented IPS'' and ``w/o item-oriented IPS'' performed much better, indicating that the propensity scores estimated from either of the two views are effective; (3) DEPS with dual propensity scores performed best, verified the effectiveness of DEPS by using both views to conduct the propensity scores estimation.

\begin{table}[t]
    \small
    \caption{Ablation study %for dual-transformer
    on Amazon-Digital-Music test set. }
    \label{tab:ablation_trans}
    %\vspace{-0.2cm}
    \centering
    \begin{tabular}{l|ccc|ccc}
        \hline
        Metric & \multicolumn{3}{c|}{NDCG@K} & \multicolumn{3}{c}{HR@K} \\
        \hline
         K  & 5 & 10 & 20 & 5 & 10 & 20 \\
        \hline
        w/o $\mathbf{h}_u^{<t}$  & 0.2161 & 0.2720  & 0.3192 & 0.3599 & 0.5467 & 0.7115 \\
        w/o $\mathbf{h}_i^{<t}$  & 0.2034 & 0.2541 & 0.3035 & 0.3819 & 0.5383 & 0.7253\\
        w/o stage-1 &  0.1953 & 0.2423  & 0.2869 & 0.3489 & 0.5082 & 0.6869 \\
        \hline
        DEPS & \textbf{0.2256} & \textbf{0.2766}  & \textbf{0.3273} & \textbf{0.4093} & \textbf{0.5852} & \textbf{0.7747}\\
        \hline
    \end{tabular}
    %\vspace{-0.2cm}
\end{table}

Dual-transformer depends on several important mechanisms for estimating propensity scores and learning model parameters, including estimating with the interaction sequences from both $\mathbf{h}_u^{<t}$'s and  $\mathbf{h}_u^{<t}$'s, and using the first stage unsupervised learning for initializing the parameters. Based on the Amazon-Digital-Music dataset, we conducted ablation studies to test the performances of DEPS variations by removing these components shown in Table~\ref{tab:ablation_trans}. 
These DEPS variations include: estimating the propensity scores and conducting recommendation without using $\mathbf{h}_u^{<t}$ (denoted as ``w/o $\mathbf{h}_u^{<t}$''), without using $\mathbf{h}_i^{<t}$ (denoted as ``w/o $\mathbf{h}_i^{<t}$''), and training by skipping the first stage tuning (denoted ``w/o stage-1'').  

\begin{table}[t]
\setlength{\tabcolsep}{4.5pt}
    \small
    \caption{Performance comparisons between DEPS and non-sequential IPS methods on Amazon-Digital-Music test set. }
    \label{tab:Ablation}
    %\vspace{-0.2cm}
    \centering
    \begin{tabular}{l|ccc|ccc}
        \hline
        Metric & \multicolumn{3}{c|}{NDCG@K} & \multicolumn{3}{c}{HR@K} \\
        \hline
         K  & 5 & 10 & 20 & 5 & 10 & 20 \\
        \hline
        %& 0.2133 & 0.2665  & 0.3156 & 0.3736 & 0.5495 & 0.7273
        % DEPS w/o propensity score & 77.73 & 0.57 \\
        Item-Pro  & 0.1984 & 0.2385  & 0.1928 & 0.3819 & 0.5220 & 0.7170 \\
        User-Pro  & 0.2044 & 0.2494 & 0.2983 & 0.3681 & 0.5275 & 0.7088\\
        Item-User-Pro &  0.2133 & 0.2665  & 0.3156 & 0.3736 & 0.5495 & 0.7273 \\
        \hline
        DEPS & \textbf{0.2256} & \textbf{0.2766}  & \textbf{0.3273} & \textbf{0.4093} & \textbf{0.5852} & \textbf{0.7747}\\
        \hline
    \end{tabular}
 %   \vspace{-0.2cm}
\end{table}

According to the results reported in Table~\ref{tab:ablation_trans}, we found that compared with the original DEPS, the performances of all DEPS variations dropped, indicating the importance of these mechanisms. Specifically, we found that the performances dropped a lot when either sequence of $\mathbf{h}_u^{<t}$'s or $\mathbf{h}_i^{<t}$'s were removed from the model, verified the importance of estimating propensity scores from both views simultaneously in sequence recommendation. The results also indicate that the first stage of unsupervised learning did enhance recommendation accuracy.

\subsubsection{Influence of Sequential IPS Estimation Methods}
In this section, we study the influence of sequential IPS estimation methods compared to non-sequential IPS estimation methods. In our model, the user-oriented and item-oriented IPS are estimated through GRU. We compared it with the non-sequential IPS estimation methods (frequency-based propensity score). The user-oriented IPS are calculated as $p_{u,*} = m_u / \max_{u'\in\mathcal{I}}m_{u'}$ and item-oriented IPS are calculated as $p_{*,i} = m_i / \max_{i'\in\mathcal{I}}m_{i'}$, where $m_u$ and $m_i$ denotes the interaction numbers of user $u$ and item $i$.

We studied their unbiased performance in the second stage of training when the unbiased loss function $\mathcal{L}^\textrm{unbiased}$ is optimized. Specifically, we showed the NDCG@K and HR@K of several DEPS variations, including learning the recommendation model with a single propensity score $p_{*,i}$ (denoted as ``Item-Pro''), with a single propensity score $p_{u,*}$ (denoted as ``User-Pro''), and with dual propensity scores $p_{*,i},p_{u,*}$ (i.e., replacing estimated propensity score $P(i,\mathbf{h}_u^{<t}),P(u,\mathbf{h}_i^{<t})$ in DEPS as $p_{*,i},p_{u,*}$ respectively, and denoted as ``Item-User-Pro'').

From the performances shown in Table~\ref{tab:Ablation}, we found that (1) ``DEPS'' outperformed ``Item-User-Pro'' by a large margin, indicating the importance of estimating the propensity scores sequentially; (2) ``Item-Pro'' and ``User-Pro'' performed worse than ``Item-User-Pro'', indicating that the propensity scores estimated from both views (user or item) are effective and complementary.

\begin{table}[t]
\setlength{\tabcolsep}{4.5pt}
    \Small
    \caption{Performance comparisons between DEPS variations where the underlying recommendation model is replaced with GRU4Rec+ or FPMC. Experiments were conducted on Amazon-Digital-Music and `$*$' means the improvements over the underlying models are statistical significant (t-test and $p$-value $< 0.05$). }
    \label{tab:model-agnostic}
    %\vspace{-0.2cm}
    \centering
    \begin{tabular}{@{}l@{}|ccc|ccc}
        \hline
        Metric & \multicolumn{3}{c|}{NDCG@K} & \multicolumn{3}{c}{HR@K} \\
        \hline
         TopK  & 5 & 10 & 20 & 5 & 10 & 20 \\
        \hline
        GRU4Rec+ & 0.0651 & 0.0961 & 0.1385 & 0.1709 & 0.3025 & 0.4874 \\
        DEPS(GRU4Rec+) & \textbf{0.0861}* & \textbf{0.1219}*  & \textbf{0.1608}* & \textbf{0.1905}* & \textbf{0.3272}* & \textbf{0.5182}*\\
        \hline
        FPMC & 0.1355 & 0.1799 & 0.2153 & 0.2941 & 0.4391 & 0.6091 \\
        DEPS(FPMC) & \textbf{0.1396}* & \textbf{0.1858}*  & \textbf{0.2189}* & \textbf{0.3025}* & \textbf{0.4566}* & \textbf{0.6426}*\\
        \hline
    \end{tabular}
    %\vspace{-0.3cm}
\end{table}

%\vspace{-0.3cm}
\subsubsection{DEPS as a Model-Agnostic Framework}
Though DEPS designs a transformer-based model for conducting the recommendation, it can also be used as a model-agnostic framework, by replacing the underlying model (the transformers and the MLP shown in Section~\ref{sec:SR_DEPS}) with other sequential recommendation models. In the experiments, we replaced it with GRU4Rec+~\cite{imporvedRec4gru} and FPMC~\cite{FPMC}, achieving two new models, denoted as ``DEPS (GRU4Rec+)'' or ``DEPS (FPMC)'', respectively. Please note that the sequential recommendation models of GRU4Rec and FPCM cannot be trained on MLM tasks. Therefore, the loss functions in the first stage training of DEPS (GRU4Rec+) and DEPS (FPMC) degenerates to $\mathcal{L}_u^{\text{AR}}+\mathcal{L}_i^{\text{AR}}$. From the results reported in Table~\ref{tab:model-agnostic}, we found that the DEPS (GRU4Rec+) and DEPS (FPMC) respectively achieved improvements over their underlying models of GRU4Rec+ and FPMC. The results indicate that the propensity scores estimated by DEPS is general. They can be used to improve other sequential recommendation models in a model-agnostic manner. 

\begin{figure}
    \centering
    \subfigure[HR@K curve]{
    \includegraphics[width=0.225\textwidth]{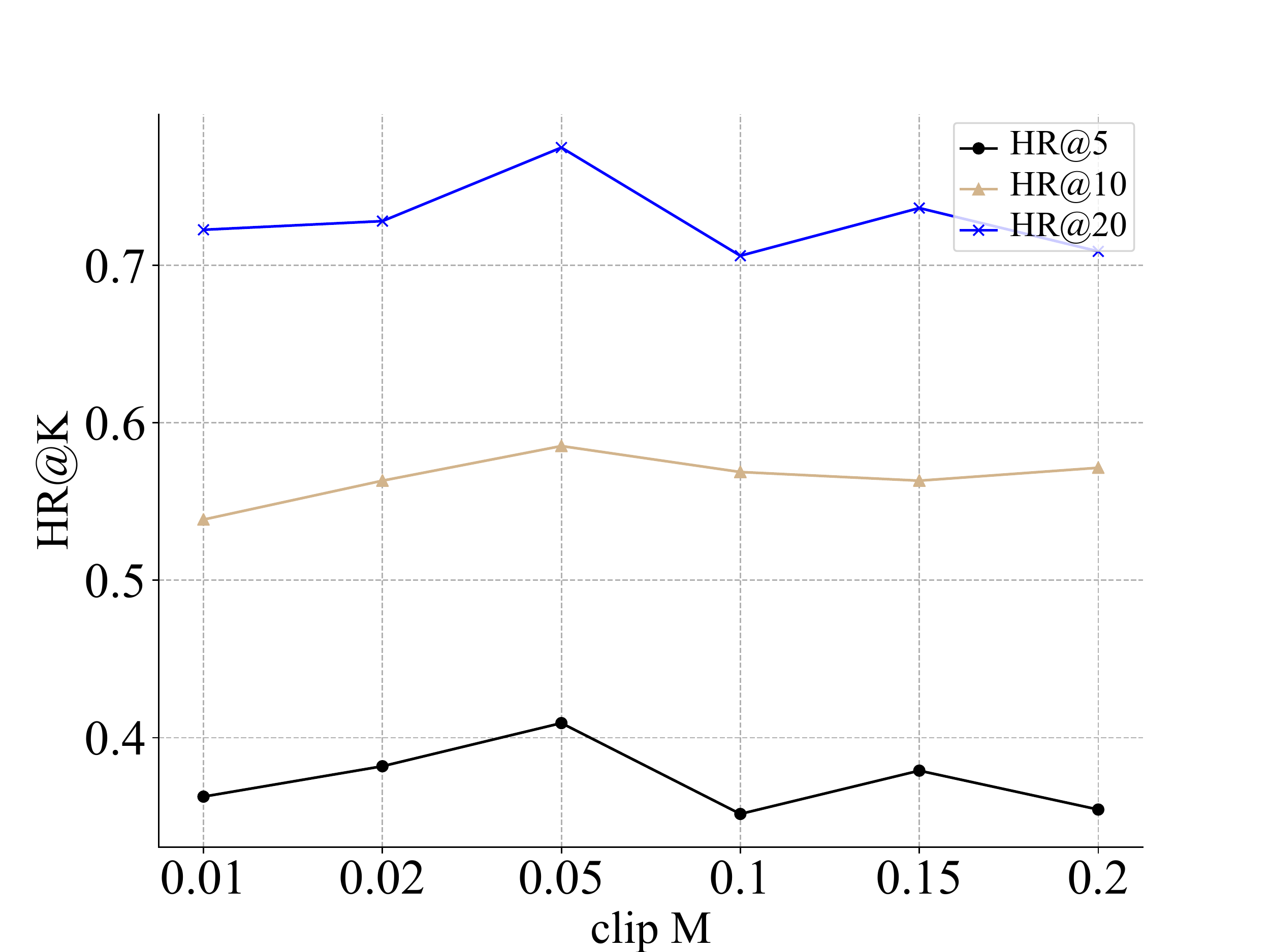}
    }
    \subfigure[NDCG@K curve]{
    \includegraphics[width=0.225\textwidth]{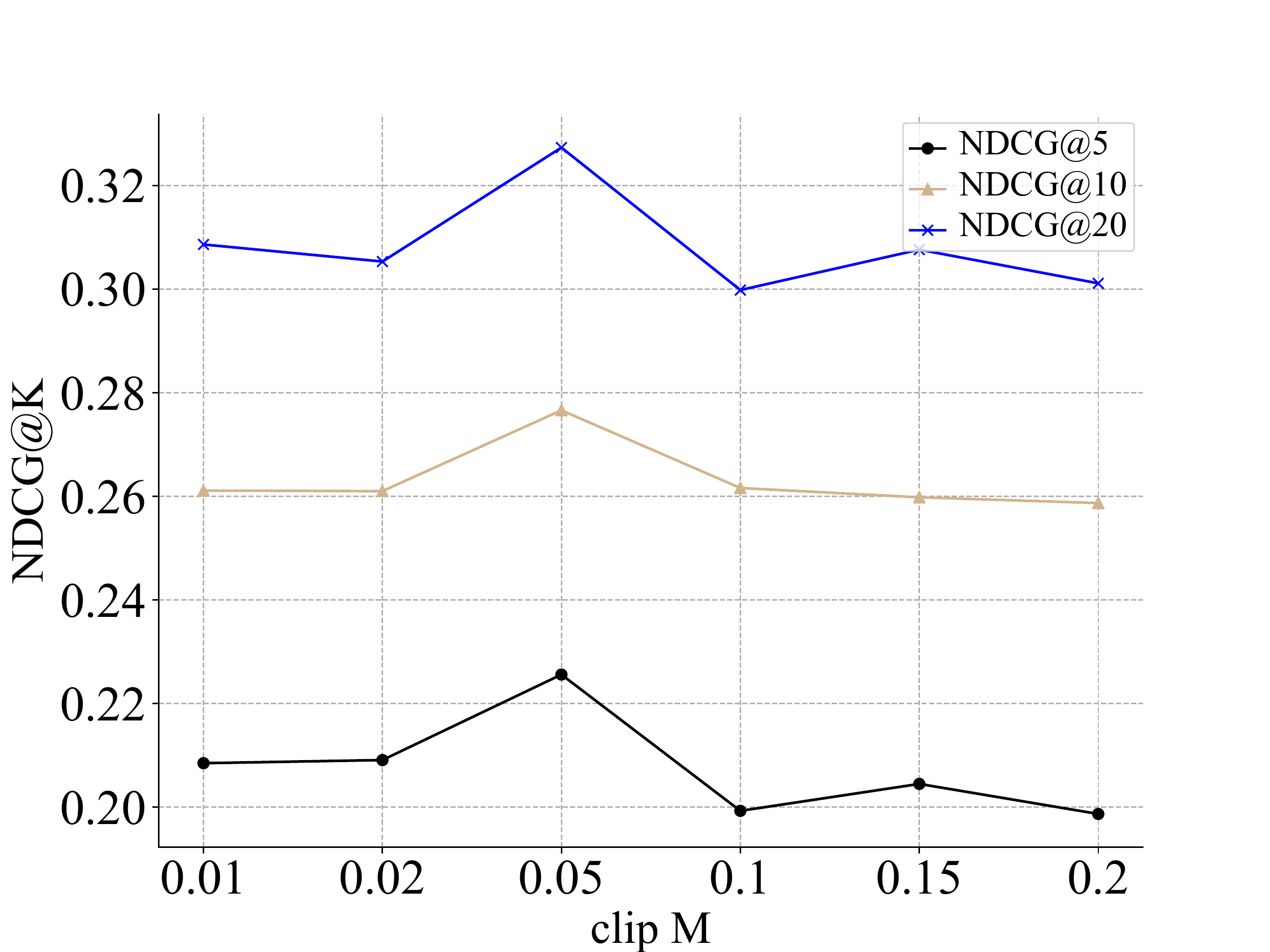}
    }
    \caption{NDCG and HR curves of DEPS w.r.t. clip value $M$}
    \label{fig:M-adjust}
\end{figure}

\subsubsection{Impact of the Clipping Value M}\label{sec:ClipValM}
According to Theorem~\ref{theo:var}, the clip value $M$ balances the unbiasedness and variance in DEPS. In this experiment, we studied how NDCG@K and HR@K changed when the clip value $M$ was set to different values from $[0.05.0.2]$. From the curves shown in Figure~\ref{fig:M-adjust}, we found the performance improved when $M\in[0.01,0.05]$ and then dropped between $[0.05,0.2]$. The results verified the theoretical analysis that too small $M$ (e.g., $M=0.01$) results in large variance estimation while too large $M$ (e.g., $M=0.2$) results in large bias. It is important to balance the unbiasedness and variance in real applications.

\section{Conclusion}
This paper proposes a novel IPS estimation method called Dually Enhanced Propensity Score Estimation (DEPS) to remedy the exposure or selection bias in the sequential recommendation. DEPS estimates the propensity scores from the views of item and user and offers several advantages: theoretical soundness, model-agnostic nature, and end2end learning. Extensive experimental results on four real datasets demonstrated that DEPS can significantly outperform the state-of-the-art baselines under the unbiased test settings.

\begin{acks}
This work was funded by the National Key R\&D Program of China (2019YFE0198200), National Natural Science Foundation of China (61872338, 62102420, 61832017), Beijing Outstanding Young Scientist Program NO. BJJWZYJH012019100020098.%, Intelligent Social Governance Interdisciplinary Platform, Major Innovation \& Planning Interdisciplinary Platform for the ``Double-First Class'' Initiative, Renmin University of China, the Fundamental Research Funds for the Central Universities and the Research Funds of Renmin University of China (No.22XNH027), and Public Policy and Decision-making Research Lab of Renmin University of China. We would like to thank Bruno L. Pereira and  Ahmed Kachkach for their valuable suggestion and help of this work.

\end{acks}

% In this paper, we proposed a novel model called Time-aware De-biasing Model (DEPS) to break the information cocoons in the sequential recommendation. In DEPS, the information cocoon is formalized as an exposure bias problem under a causal view, which can be addressed by estimating the propensity scores based on interaction sequences derived from both view of user and the view of item. DEPS offers several advantages: theoretical soundness, model-agnostic nature, and end2end learning. Extensive experimental results on four publicly available datasets demonstrated that DEPS can not only significantly outperform the state-of-the-art baselines but also bring more item diversity to users.
\appendix
%\section*{Appendix}\label{sec:Appendix}
%\begin{theorem}
% Theorem~\ref{theo:objective} 
% \textit{The objective function $\mathcal{L}^{unbiased}$ is an unbiased estimation of the training objective $\mathcal{L}^{ideal}$.}
% % \end{theorem}

% \begin{proof}
% \begin{align*}
%     \mathcal{L}^{unbiased} &= \mathbb{E}_{o}\left[\mathbb{E}_{u,i,t\sim\mathcal{D}}\left[\frac{\delta(c,\hat{r})}{p(o=1)}\right]\right] \\
%     &= \mathbb{E}_{u,i,t\sim\mathcal{D}}\left[\mathbb{E}_{o}\left[\frac{\delta(r,\hat{r})o}{p(o=1)}\right]\right]\\
%     &= \mathbb{E}_{u,i,t\sim\mathcal{D}}\left[\delta(r,\hat{r}(u,i,\mathbf{h}^{<t}_i,\mathbf{h}^{<t}_u))\right] =  \mathcal{L}^{ideal}
% \end{align*}
% \end{proof}
\section{Proof of Theorems}

\subsection{Proof of Theorem~\ref{theo:ProbEstiamtion}}\label{sec:ProofTheorem1}

%\textit{Both of the propensity score estimations $p(u|\mathbf{h}^{<t}_i)$ and $p(i|\mathbf{h}^{<t}_u)$ are the unbiased estimations of propensity score, which also denotes that $\mathcal{L}^{unbiased}  = (\mathcal{L}^{unbiased}_i+\mathcal{L}^{unbiased}_u)/2$.}
% \end{theorem}

\begin{proof}
%According to our time-aware causal graph shown in Figure~\ref{fig:CausalGraph}, we assume that the item $i$ will be observed $o=1$ if it appeared in the historical interaction sentence, i.e. $o = q_{\mathbf{h}^{<t}_u,i}$.
Let $q_{\mathbf{h}^{<t}_u,i} \in \{0,1\}$ where $q_{\mathbf{h}^{<t}_u,i} = 1$ indicates item $i$ is observed in the historical interaction sentence $\mathbf{h}^{<t}_u$, otherwise $q_{\mathbf{h}^{<t}_u,i} = 0$.    According to the definition, $P(q_{\mathbf{h}^{<t}_u,i}=1) = P(i,\mathbf{h}^{<t}_u)$ in sequential recommendation. Abbreviate historical information $\mathbf{h}_u^{<t},\mathbf{h}_i^{<t}$ to $H$. Therefore,
\begin{align*}
    \mathbb{E}_o\left[\mathcal{L}_u\right] &= \mathbb{E}_o\left[\sum_{u\in\mathcal{U}}\sum_{(i,c_t)\in\mathcal{D}^u}\left[\frac{\delta(c_t,\hat{r}_t(u,i^t,H))}{P(i^t,\mathbf{h}^{<t}_u)}\right]\right] \\
     &= \sum_{u\in\mathcal{U}}\sum_{t:(u,i', c_t)\in\mathcal{D}}\sum_{i\in\mathcal{I}}\mathbb{E}_{q}\left[q_{\mathbf{h}^{<t}_u,i}\cdot\frac{\delta(r_t,\hat{r}_t(u,i,H))}{P(i,\mathbf{h}^{<t}_u)}\right] \\
    &= \sum_{u\in\mathcal{U}}\sum_{t:(u,i', c_t)\in\mathcal{D}}\sum_{i\in\mathcal{I}}\delta(r_t,\hat{r}_t(u,i,H)) = \mathcal{L}^{\text{ideal}}_s \\
\end{align*}
% \begin{align*}
%     \mathcal{L}_u &= \mathbb{E}_{o}\left[\mathbb{E}_{(u,i,t,c)\sim\mathcal{D}}\left[\frac{\delta(c,\hat{r})}{P(i|\mathbf{h}^{<t}_u)}\right] \right] = \mathbb{E}_{q}\left[\mathbb{E}_{(u,i,t,c)\sim\mathcal{D}}\left[\frac{\delta(c,\hat{r})}{P(i|\mathbf{h}^{<t}_u)}\right] \right] \\
%     &= \mathbb{E}_{(u,i,t,c)\sim\mathcal{D}}\left[\mathbb{E}_{q_{\mathbf{h}^{<t}_u,i}}\left[\frac{\delta(r,\hat{r})}{P(i|\mathbf{h}^{<t}_u)}q_{\mathbf{h}^{<t}_u,i}\right] \right] \\
%     &= \mathbb{E}_{(u,i,t,c)\sim\mathcal{D}}\left[\delta(r,\hat{r}(u,i,\mathbf{h}^{<t}_i,\mathbf{h}^{<t}_u))\right] =  \mathcal{L}^\text{ideal}.
% \end{align*}

Similarly, let $q_{\mathbf{h}^{<t}_i,u} \in \{0,1\}$ where $q_{\mathbf{h}^{<t}_i,u} = 1$ indicates the user $u$ appeared in the historical interaction sentence $\mathbf{h}^{<t}_i$. We have $P(q_{\mathbf{h}^{<t}_i,u}=1) = P(u,\mathbf{h}^{<t}_i)$, and
\begin{align*}
    \mathbb{E}_o\left[\mathcal{L}_i\right] &= \mathbb{E}_o\left[\sum_{i\in\mathcal{I}}\sum_{(i,c_t)\in\mathcal{D}^i}\left[\frac{\delta(c_t,\hat{r}_t(u^{l(i,t+1)},i,H))}{P(u^{l(i,t+1)},\mathbf{h}^{<t}_i)}\right]\right] \\
     &= \sum_{i\in\mathcal{I}}\sum_{t:(u,i', c_t)\in\mathcal{D}}\sum_{u\in\mathcal{U}}\mathbb{E}_q\left[q_{\mathbf{h}^{<t}_i,u}\frac{\delta(r,\hat{r}(u,i,H))}{P(u,\mathbf{h}^{<t}_i)}\right] \\
    &= \sum_{u\in\mathcal{U}}\sum_{t:(u,i', c_t)\in\mathcal{D}}\sum_{i\in\mathcal{I}}\delta(r_t,\hat{r}_t(u,i,H)) = \mathcal{L}^{\text{ideal}}_s.
\end{align*}
Therefore,
$\mathcal{L}^\mathbf{ideal}_s = \mathbb{E}\left[\mathcal{L}^\mathbf{unbiased}_s\right] = \mathbb{E}\left[\alpha\mathcal{L}_u+(1-\alpha)\mathcal{L}_i\right]$
\end{proof}

\subsection{Proof of Theorem~\ref{theo:var}}\label{sec:ProofTheorem2} 

%Theorem~\ref{theo:var} \textit{Let $L_{u}^t,L_{i}^t$ be the random variable of the single sample loss, i.e. $L_{u}^t = \frac{\delta(c,\hat{r})}{P(u|\mathbf{h}^{<t}_i)},L_{i}^t = \frac{\delta(c,\hat{r})}{P(i|\mathbf{h}^{<t}_u)}$,
%we have
%$$\mathbb{V}\left[\frac{1}{2}(L_{u}^t + L_{i}^t)\right] \leq \\
%\frac{1}{2}\max\left\{\mathbb{V}\left[L_{u}^t\right],\mathbb{V}\left[L_{i}^t\right]\right\},$$
%where $\mathbb{V}[\cdot]$ denotes the estimation variance.}

\begin{proof}
Following the notations defined  Theorem~\ref{theo:ProbEstiamtion}, we can write $L_{u}^t=\frac{\delta(r_t,\hat{r}_t)}{p(u,\mathbf{h}^{<t}_i)}q_{\mathbf{h}^{<t}_i,u}$, and according to the clip operation shown in Eq.~(\ref{eq:clip_prob_user}) we have $\widetilde{P}(u,\mathbf{h}^{<t}_i) \ge M$. Therefore,
\begin{align*}
    \mathbb{V}\left[L_{u}^t\right] &= \mathbb{E}\left[L_{u}^t\right]^2-\left(\mathbb{E}\left[L_{u}^t\right]\right)^2 = \frac{\delta^2(r_t,\hat{r}_t)}{\widetilde{P}(u,\mathbf{h}^{<t}_i)}-\delta^2(r_t,\hat{r}_t)\\
    &= \left(\frac{1}{\widetilde{P}(u,\mathbf{h}^{<t}_i)}-1\right)\delta^2(r_t,\hat{r}_t) \leq \left(\frac{1}{M}-1)\delta^2(r_t,\hat{r}_t)\right).
\end{align*}
Similarly, we can bound the  variance of objective functions $L_{i}^t$ as,
\[
\mathbb{V}\left[L_{i}^t\right] =\left (\frac{1}{\widetilde{P}(i,\mathbf{h}^{<t}_u)}-1\right)\delta^2(r_t,\hat{r}_t) \leq \left(\frac{1}{M}-1\right)\delta^2(r_t,\hat{r}_t).
\]
Applying the Cauchy-Schwarz' inequality, we can bound the variance of average loss $\mathcal{L}_s^\textrm{unbiased}=\alpha L_{i}^t+(1-\alpha)L_{u}^t$ as:
\begin{small}
\begin{align*}
    \mathbb{V}\left[\mathcal{L}_s^\textrm{unbiased}\right]
    &= \alpha^2\mathbb{V}\left[L_{i}^t\right] + (1-\alpha)^2\mathbb{V}\left[L_{u}^t\right]+ 2\alpha(1-\alpha)\text{Cov}(L_{i}^t,L_{u}^t) \\
    &\leq \alpha^2\mathbb{V}\left[L_{i}^t\right] + (1-\alpha)^2\mathbb{V}\left[L_{u}^t\right]+ 2\alpha(1-\alpha)\sqrt{\mathbb{V}\left[L_{i}^t\right]\mathbb{V}\left[L_{u}^t\right]}\\
    &\leq \max\left\{\mathbb{V}\left[L_{i}^t\right],\mathbb{V}\left[L_{u}^t\right]\right\} \leq (\frac{1}{M}-1)\delta^2(r,\hat{r}).
\end{align*}
\end{small}
\end{proof}
\bibliographystyle{ACM-Reference-Format}
\balance
\bibliography{ref}

%%
%% If your work has an appendix, this is the place to put it.
%% \appendix

\end{sloppy}
\end{document}